\documentstyle[aps,multicol]{revtex}\draft

\def\beginwide{
        \end{multicols} \vspace*{-0.5cm} \noindent
        \rule{3.5in}{.1mm}\rule{.1mm}{5mm} \widetext \medskip }
\def\beginwidetop{
        \end{multicols} \vspace*{-0.5cm} \noindent
        \widetext \medskip }
\def\endwide{
        \hspace*{3.35in}~\rule[-5mm]{.1mm}{5mm}\rule{3.5in}{.1mm}
        \begin{multicols}{2} \vspace*{-1.0cm} \noindent }
\def\endwidebottom{
        \begin{multicols}{2} \vspace*{-1.0cm} \noindent }

\newcommand{\beq}{\begin{equation}}
\newcommand{\eeq}{\end{equation}}
\newcommand{\bdis}{\begin{displaymath}}
\newcommand{\edis}{\end{displaymath}}
\newcommand{\bea}{\begin{eqnarray}}
\newcommand{\eea}{\end{eqnarray}}
\newcommand{\barr}{\begin{array}}
\newcommand{\earr}{\end{array}}
\begin{document}

\title{Dynamically Driven Renormalization Group Applied to
Sandpile Models}
\author{
Eugene V. Ivashkevich$^{(1,2)}$,
Alexander M. Povolotsky$^{(2)}$,\\
Alessandro Vespignani$^{(3)}$,
and Stefano Zapperi$^{(4)}$}

\address{
1)Dublin Institute for Advanced Studies,
10 Burlington Road, Dublin 4, Ireland\\
2)Bogoliubov Laboratory of Theoretical
Physics, J.I.N.R., Dubna 141980, Russia\\
3)The Abdus Salam International Centre for Theoretical Physics (ICTP),
P.O.Box 586, 34100 Trieste, Italy\\
4)Center for Polymer Studies and Department of Physics
Boston University, Boston, MA 02215}
\date{\today}
\maketitle
\begin{abstract}
The general framework for the renormalization group analysis of
self-organized critical sandpile
models is formulated. The usual real space renormalization
scheme for  lattice models when
applied to nonequilibrium dynamical models must be supplemented by
feedback relations coming from the stationarity  conditions. On the
basis of these ideas the Dynamically Driven Renormalization Group  is
applied to describe the  boundary and bulk
critical behavior of  sandpile models. A detailed description of the
branching nature of sandpile avalanches is given in terms of the
generating functions of the underlying branching process.
\end{abstract}

\begin{multicols}{2}
\section{Introduction}

In the last decade it has been recognized that a fair amount
of physical phenomena are characterized  by
strong fluctuations and
long-range correlation functions.
According to the theory of equilibrium statistical physics,
we expect scale invariance only in presence
of certain symmetries  or at critical points \cite{domb}
We are therefore lead to seek for the origin of the scale invariance
in nature in the rich domain of nonequilibrium
systems \cite{katz,zia,vic,bb}.
One might hope, in fact, that there are classes of nonequilibrium systems
that generate scale invariance for a wide (and arbitrary) range
of physical parameter, providing an explanation for the commonly
observed scaling laws.

Pursuing this aim, Bak, Tang and Wiesenfeld proposed the concept of
Self-Organized Criticality (SOC) \cite{soc} as a unifying framework to
describe a vast class of dynamically driven systems  which evolve
spontaneously  in a stationary state with a broad power law
distribution of discrete energy dissipating events.
To illustrate the basic ideas of SOC, they
introduced a cellular automaton  model of sandpiles.
In this model, criticality seems to emerge automatically
if the system is driven at an infinitesimal rate\cite{bak2,bak,ds}.
Because of the enormous conceptual potentiality,
SOC ideas have reverberated rapidly throughout the sciences, from
geophysics to economics and biology, as a prototype mechanism to
understand the manifestation of scale invariance and complexity in
natural phenomena.

The major source of difficulties in the study of
sandpiles models is
the absence of a general criterion,
like the use of the Gibbs distribution in equilibrium systems,
to assign an ensemble statistical measure to a particular
configuration of the system. This problem is common to many
nonequilibrium systems whose theoretical
understanding lies  far behind the equilibrium theory.
In particular, 
many relations among sandpiles automata and systems 
with nonequilibrium absorbing critical point have been 
recently enlightened \cite{vzmf}.

In the past years, we developed a renormalization group (RG) strategy
for sandpile models \cite{pvz}
which  has also  been  applied \cite{lpvz} to forest-fire
models \cite{bak,ds}.
This approach deals with the critical properties of
the system by introducing in the renormalization equations a dynamical
steady state condition
which provides non-equilibrium stationary statistical weights to be
used in the calculation.
This scheme, named the Dynamically Driven Renormalization Group
(DDRG) \cite{ddrg}, has been successively generalized as a  renormalization
framework for systems with a non-equilibrium critical steady-state.
Recently, the DDRG approach  has been improved
including higher order proliferations
through a general scheme \cite{iva}. The method has also been applied
to one dimensional sandpiles \cite{wies} and
other non-equilibrium systems \cite{oliv}.

Here we discuss the application of the DDRG to sandpile models,
deriving systematically the  previous RG schemes \cite{pvz,iva}
and presenting new results.
We will introduce the general strategy of the DDRG for the critical height
sandpile models, and its practical implementation for increasingly complex
proliferation schemes. In order to treat such a high level of calculation
complexity, we introduce a generating function for
the basic recursion relations.
The scheme is then extended by exploiting the analogy with a particular
chain chemical reaction.
Finally its application to the calculation of the boundary critical behavior
is shown.

The paper is organized as follows: in Sec.2 we introduce the
 class of sandpile  automata and its mapping into a general
nonequilibrium cellular automaton (CA). Sec.3 presents the
Dynamically Driven Renormalization Group general scheme.
Sec.4 shows the explicit application of the DDRG to the sandpile
in its simple scheme. In Sec. 5 we present the actual calculations of the
renormalization equations and their generating function
and results obtained.
Sec.6 describes the extended chemical reaction scheme and its
results. Sec. 7 is devoted to the renormalization analysis of
the boundary critical behavior. Sec. 8 presents the summary and
conclusions.

\section{The sandpile model}

The prototype example  for SOC is provided by sandpiles:
sand is added grain by grain
until unstable sand (too large local slope of the
pile) slides off. In this way the pile reaches a steady-state,
in which additional sand grains fall
off the pile by avalanche events.
The steady-state is critical since avalanches of any size are observed.
This class of models can be used to describe a generic avalanche
phenomenon, interpreting the sand as energy, mechanical stress or heat memory.

Sandpile models are cellular automata \cite{bak2,zhang}
defined in a $d-$dimensional
lattice. A discrete or continuous variable $E(i)$,
that we denote by energy, is
associated with each lattice site $i$. At each time step
an input energy $\delta E$ is added to  a randomly chosen site.
When the energy on a site reaches a threshold value $E_c$, the
site relaxes transferring energy to the neighboring sites
\beq
E(i)\to E(i)-\sum_{e}\Delta E(e) \nonumber
\eeq
\beq
E(i+e)\to E(i+e)+\Delta E(e)\label{din}
\eeq
where $e$ represents the unit vectors on the lattice. A typical choice
for the parameters is , for example, $E_c=4$ and $\Delta E(e)=\delta E=1$,
but other possibilities have also been considered.
The relaxation of the first site can induce a series of relaxations
generating an avalanche. Note that the energy is added to
the system only when the configuration is stable (i.e. all the
sites are below the threshold).
The boundary conditions are usually chosen to be
open so that energy can leave the system.
In these conditions the system organizes itself into a stationary
state characterized by avalanches of all length scales.
In particular, the distribution for avalanches sizes $s$ decays
as a power law $P(s)\sim s^{-\tau}$, and the linear size of
the avalanche scales with time $r\sim t^z$.
This model has been extensively studied in the past by means of
numerical simulations \cite{manna,grasma,stella,lubeck} and
several exact results have been derived for
abelian sandpiles models (ASM)\cite{D,MDheight,P,I,IKP}.

Given the above definition of sandpiles we can rephrase them in
the language of discrete nonequilibrium probabilistic CA.
To each site $i$
is associated a variable $s_i$ that can assume
$q$ different values ($s_i=1, 2, 3,\cdots, q$). For
instance, each state might correspond to an allowed  energy
level. The subscript $i$ labels the lattice site.
 A complete set $s\equiv\{s_i\}$ of
lattice variables specifies a configuration of the system.
We define
$\langle s\mid T(\mu) \mid s^0\rangle$
as the transition rate
from a configuration $s_0$ to
a configuration $s$ in a time step $t$
as a function of a set of parameters  $\mu=\{\mu_i\}$.
SOC automata are usually defined by a transition probability
given by the product
\beq
\langle s\mid T \mid s^0\rangle=
\prod_{i=1}^{N} \tau(s_i\mid s_i^0,\{s_{i+e}^0\})
\eeq
where $N=L^d$ is the number of sites, and $e$ specifies the nearest neighbor
(n.n.) vector. The dynamics is therefore expressed
as a product of one-site transition probabilities depending upon
the sites and its nearest-neighbors states at the previous time step.

As we said, the common characteristic of SOC systems is the presence of
a nonequilibrium critical steady-state, which
we can analyze using the DDRG formalism.
However, it is worth remarking that SOC systems reach true criticality
just in the limit of an infinite slow driving condition. This means that
the perturbing time scale is much larger than the dynamical activity one.
SOC systems relax far more rapidly than they are perturbed. In
practice, this implies that no new grain of sand is dropped until the avalanches
started by the previous grain has finished. In this way avalanches cannot
overlap, and their dynamics is well defined with respect to the external
field. A complete RG analysis should take into account also the driving field.
However, since we are interested in the critical point, we will study
the system in the limit of slow driving. A more detailed discussion of the 
complete sandpile automaton phase diagram is provided in ref.\cite{vzmf}.

\section{The DDRG}

The probability distribution of CA such as those shown in the previous
section obeys the following master equation (ME)
\beq
P(s,t_0+t)=\sum_{\{s^0\}} \langle s\mid
T(\mu) \mid s^0\rangle
P(s^0,t_0).
\eeq
The explicit solution of the master equation is not in general
available but we can extract the critical properties
of the model by a renormalization group analysis.
We coarse grain the system by rescaling lengths and time according to
the transformation $x\to b^{-1}x$ and $t\to b^{-z}t$.
The renormalization transformation is
constructed through the operator
${\cal R} (S,s)$ that
introduces a set of coarse grained variables $S\equiv\{S_i\}$
and rescales the lengths of the system \cite{nv}.
In general, $\cal R$ is a projection operator with the properties
${\cal R}(S,s)\geq 0$ for any $\{S_i\},\{s_i\}$, and
$\sum_{\{S\}} {\cal R}(S,s)=1$.
These properties preserve the normalization condition
of the renormalized distribution. The explicit form of the operator $\cal R$
is defined case by case in various applications of the method.
Usually, it corresponds to a block transformation in which lattice
sites are grouped together in a super-site
that  defines the renormalized
variables $S_i$ by means of a majority or a spanning rule.

We subdivide the time step in intervals of the unitary time scale ($t_0=0$)
obtaining the coarse graining of the system as follows:
\bea
\lefteqn{P^{\prime}(S,t')= }\nonumber \\
&\sum_{\{s\}}{\cal R}(S,s)
\sum_{\{s^0\}}\langle s\mid T^{b^z}(\mu)\mid s^0\rangle
P(s^0,0)
\label{start}
\eea
where we have included the application of the operator
$\cal R$ and $t'=b^z t$. The meaning
of $\langle s\mid T^{b^z}(\mu) \mid s^0\rangle$
has to be defined explicitly:
the simplest possibility is $b^z=N$ where  $N$ is an integer number,
and $T^N$ denotes the application of the dynamical operator $N$ times.
In general, since we are dealing with a discrete time
evolution, we have to consider $T^{b^z}$ as a convolution over
different paths, chosen by an appropriate condition.
The detailed definition of the effective operator $T^{b^z}$
for the sandpile is reported in the next section.
By multiplying and dividing each term of eq.~(\ref{start})
by $P^{\prime}(S^0,0)=\sum_{\{s^0\}}{\cal R}(S^0,s^0)
P(s^0,0)$ and using the properties of the operator ${\cal R}$,
after some algebra we get
\beginwide
\beq
P^{\prime}(S,t')=
\sum_{\{S^0\}}(\frac
{\sum_{\{s^0\}}\sum_{\{s\}}
{\cal R}(S^0,s^0)
{\cal R}(S,s)\langle s\mid T^{b^z}(\mu) \mid s^0\rangle
P(s^0,0)}
{\sum_{\{s^0\}}{\cal R}(S^0,s^0)
P(s^0,0)})P^{\prime}(S^0,0)
\label{mefine}
\eeq
\endwide
which finally identifies  the renormalized dynamical operator
$\langle S\mid T'\mid S^0\rangle$. In other words, the new dynamical
operator $T'$ is the sum over all the dynamical paths of $b^z$ steps
that from a starting configuration $\{s^0_i\}$ lead to a configuration
$\{s_i\}$ which renormalizes respectively in $\{S^0_i\}$ and $\{S_i\}$.
The sum is weighted by the normalized
statistical distribution of each configuration.
The scheme discussed so far is a general formulation valid
for each system which
exhibits a stationary state, and its application
presupposes the knowledge of the
explicit form of the steady-state
distribution $W(s)=P(s, t\to\infty)$. For instance,
in equilibrium phenomena $W(s)$ is given by the Gibbs distribution.
In this case it is possible
to apply several methods such as cumulant expansions and  exact or approximate
decimation to obtain the form of the recursion relations.
For non-equilibrium dynamical systems, in general we do not know the
form of the steady-state distribution.
We will therefore develop an approximate method to evaluate
the stationary distribution to be used in the calculation of
the renormalized master equation.

The steady-state distribution can in general be split into two parts
\beq
W(s)=W^{\mbox{(i)}}(s)+W^{\mbox{(c)}}(s)
\label{decom}
\eeq
where $W^{\mbox{(i)}}(s)$ and $W^{\mbox{(c)}}(s)$ are,
respectively, the incoherent and coherent part of the distribution.
The incoherent part of the distribution does not include
correlations among variables
and expresses a mean field approximation for the system.
The coherent part $W^{\mbox{(c)}}(s)$ can be
subdivided in parts describing
different kinds of correlations: nearest-neighbors,
next-nearest-neighbors,
etc. The incoherent part is a factorized distribution
that, for systems characterized by a q-state variables
(see sect.2), has the form
\beq
W^{\mbox{(i)}}(s)= \prod_{i}\langle\rho_{s_i}\rangle
\label{produ}
\eeq
where $\langle\rho_\kappa\rangle$ is the average density of sites in the
$\kappa$-state. In this way, we have approximated the
probability of each configuration
$\{s_i\}$ as a product measure of the mean field  probability to
have a state $s_i$ in each corresponding site.
The incoherent part contribution to the renormalization equation
can be obtained by  stationarity conditions for the system
${\cal S}_\mu (\{\langle\rho_\kappa\rangle\})=0$
to evaluate the densities $\langle\rho_\kappa\rangle$.
These conditions are derived from dynamical mean field equations
which describe the {\em driving} of the system to the nonequilibrium
steady state by means of balance constraints. The operator
${\cal S}_\mu$ depends upon the same dynamical parameters of the
operator $T$, and
by solving the stationary condition equation,
the average densities of the $\kappa$-states for the coarse grained system
are obtained as a function of $\mu$ at the corresponding
iteration of the RG equations.
By inserting the approximate distribution
in Eq.~(\ref{mefine}), we thus get the following set of renormalization equations
\beginwide
\beq
\langle S\mid T'(\mu) \mid S^0\rangle=
\frac{\sum_{\{s^0\}}\sum_{\{s\}}
{\cal R}(S^0,s^0)
{\cal R}(S,s)\langle s\mid T^{b^z}(\mu) \mid s^0\rangle
\prod_{i}\langle\rho_{s_i^0}\rangle}
{\sum_{\{s^0\}}{\cal R}(S^0,s^0)
\prod_{i}\langle\rho_{s_i^0}\rangle}
\label{ddrg1}
\eeq
\endwide
\beq
{\cal S}_{\mu}(\{\langle\rho_\kappa\rangle\})=0
\label{ddrg2}
\eeq
where the second equation denotes the dynamical steady state condition
that allows evaluation of
the approximate stationary distribution
at each coarse graining scale.
We call Eq.~(\ref{ddrg2}) the {\em driving  condition},
since it drives
the RG equations acting as a feedback on the scale transformation.
Eqs~(\ref{ddrg1}) and (\ref{ddrg2}) are the basic renormalization equations
from which the desired recursion relations are derived.
Imposing that the renormalized
operator $T'$ has the same functional form of the operator $T$,
i.e. $T'(\mu)=T(\mu')$, we obtain the rescaled
parameter set $\mu'=f(\mu)$. This implies that the renormalized  single time
distribution $P^{\prime}(S,t')$ has the same functional form of the original
distribution $P(s,t)$. The critical behavior of the model
is obtained by studying the fixed points $\mu^*=f(\mu^*)$.
Since  we are dealing with discrete evolution operators $T$, we define
the time scaling factor $b^z$ as the average number of steps and
apply the operator $T$ in order to obtain that $T'(\mu)=T(\mu')$ for the
coarse grained system.
In this way we obtain a time recursion
relation $t'=g(\mu)t$, or equivalently $b^z=g(\mu)$,
 from which it is possible
to calculate the dynamical critical exponent
$z={\log g(\mu^*)}/{\log b}$.
In this form of the DDRG, we take into account only the uncorrelated part
of the steady-state probability distribution. The results obtained
are not trivial because correlations in the systems are considered in
the dynamical renormalization of the operator $T$, that given a starting
configuration traces all the possible paths leading to the renormalized
final configuration. Moreover, geometrical correlations are
treated by the operator $\cal R$ that maps the system by means of
spanning conditions or majority rules.
The renormalized uncorrelated part of the stationary distribution is
evaluated from the stationary condition with renormalized parameters,
thus providing an effective treatment of correlations.
One can then improve the results by including higher order
contributions to the unknown stationary distribution $W(s)$
using cluster variation methods \cite{dick}.

\section{Renormalization scheme for sandpile models}

Here, we show in detail the DDRG scheme for sandpile models. For the sake of
clarity, we start by considering
the minimum proliferation
scheme. A more refined scheme is
discussed in the following sections.

To simplify the description of
sandpile models as much as possible,
we can reduce the number of states of each site
in the following way. At any scale, we divide the sites in {\em critical}
($s_i=1$) and {\em stable} ($s_i=0)$.
Stable sites do not relax
when energy is added to them. On the other hand, critical sites
relax when they receive an energy grain $\delta E_{in}$.
In this formalism we define $<\rho>$ as the density of critical sites.
For convenience, we also define unstable sites ($s_i=2)$ , as those
that are relaxing, even though they are not present in the
static configurations of the system.
These definitions can be
extended to a generic scale $b$. For instance, a cell at scale $b$ is
considered critical if the addition of energy $\delta E_{in}(b)$ induces a
relaxation of the size of the cell (i.e. the avalanche
spans the cell).

In a relaxation event at the minimal scale energy is equally distributed in the
four directions. This is no longer the case at a coarse grained
level where different possibilities arise:
the energy in principle can be distributed to one, two, three
or four neighbors. It is also worth remarking that in a certain
case unstable sites at the coarse grained scale  do not dissipate
energy to nearest neighbors, representing just intra-site energy
rearrangements. These processes define the probability that
relaxation events take place on the renormalized scale without
energy transfer. All these events occur with probabilities
$$\vec{P}=(p_0,p_1,p_2,p_3,p_4)$$
In terms of the matrix element $\langle 0|T|2\rangle$
the vector $\vec{P}$ represents the probabilities
\beq
p_n=\langle 0|T|2\rangle_n
\eeq
where $\langle 0|T|2\rangle_n$ is the probability
that a relaxing site  becomes stable and transfers
energy to $n$ neighbors.
In this way, we have obtained the set of parameters  that
describes the dynamics. Of course, the choice
of the parameter space is not uniquely determined; one encounters
proliferation problems typical of real space RG methods.
For instance, higher order proliferations are
due to multiple relaxations
of the same site and sites becoming critical during the
dynamical process (i.e.: $\langle 1|T|2\rangle$). In the following,
the practical implementation of the method considers just the
minimal proliferation we have reported above. In the next sections, a more
refined scheme will be treated.

First of all, let us show how the
driving condition is obtained by imposing the stationarity of the process.
The average  energy of a site evolves according to the following
equation written in the continuum notation:
\beq
\frac{dE(t)}{dt}=\delta E_{in}-\delta E_{out}
\label{energy-drive}
\eeq
where $\delta E_{in}$ is the average energy entering into the site
either because of relaxation in a neighboring site or
because of the external perturbations, and  $\delta E_{out}$
is the average energy dissipated by the site.
The stationary state is
characterized by the balance between the energy that goes in and the
energy that goes out of the system. We assume that energy is transferred
in ``quanta'' $\delta E = \delta E_{in}$ in each direction and we
on average obtain
\beq
\delta E =\langle\rho\rangle\delta E\sum_n np_n
\eeq
which implies
\beq
\langle\rho\rangle=\frac{1}{\sum_n np_n}
\label{balance}
\eeq
This relation gives the average density of critical
sites in the steady-state,
allowing us to evaluate the approximate stationary
distribution at each scale.

The renormalized matrix element
is then obtained by considering all the renormalized  processes that span
the cell
and transfer energy outside
\beq
p'_n = \langle S_i=0\mid T'
\mid S_i^0=2\rangle_n.
\label{sandren}
\eeq
We proceed in defining explicitly a renormalization procedure for the
dynamics by considering a finite truncation on four-sites cells.
This corresponds to a cell-to-site transformation on a square lattice,
in which each cell at the coarser scale is formed by four
sub-cells at the finer scale: the length scaling factor is $b=2$.
In this case, the operator ${\cal R}$ can be written in the following way:
\beq
{\cal R}(S,s)=\prod_J{\cal R}(S_J, \{s_i\}_J)
\eeq
where each term is acting on a specific cell $J$ and $\{s_i\}_J$
denotes the configurations of sites belonging to that cell.
A cell is renormalized
as a relaxing one if it contains a relaxing sub-cell which transfers
energy to a critical sub-cell. In this way, we ensure that the occurring
relaxation process is extending over the size of the renormalized length scale
independently of the successive avalanche evolution. A critical cell is
therefore defined by a cell which can be spanned by a path of relaxation
events. The scheme
considers only connected paths that span the cell
from left to right or top to bottom. This spanning rule implies that only
paths extending over the size of the resulting length scale contribute
to the renormalized dynamics, and it ensures the connectivity
properties of the avalanche in the renormalization procedure.

Every cell at the coarser scale can be characterized by an index $\alpha$
that indicates the configuration  of  sub-cells, and we have that
$\sum_{\{s_i\}}\to\sum_{\alpha}$.
The approximated stationary distribution (Eq.~(\ref{produ})) for
each of these configurations is given by
\beq
W_\alpha(\langle\rho\rangle)=
n_{\alpha}\prod_{i=1}^4\langle\rho_{s_i}\rangle
\eeq
where $n_{\alpha}$ is a factor due to the multiplicity of each configuration.

By using this scheme and replacing  sums
over configurations with sums over the index $\alpha$, the recursion
relations can then be rewritten in the simpler
form
\beq
p'_n=\frac{1}{\cal N}\sum_{\alpha}
W_{\alpha}(\langle\rho\rangle)
\sum_{\alpha'}\langle\alpha'\mid T^{b^z}(p_{n'}) \mid \alpha\rangle _n
\label{recsand}
\eeq
where $\mid \alpha\rangle$, $\mid \alpha'\rangle$ denotes the
four site configurations which renormalize
in $\mid S_i^0=2\rangle$ and  $\mid S_i=0\rangle$, respectively.
In the above expression the denominator of eq.(\ref{ddrg1})
is adsorbed in the normalization factor ${\cal N}$.

The effective operator  $T^{b^z}$
contains all the dynamical processes that contribute
to the definition of a meaningful renormalized
dynamics. We define the following transformation
\beq
\langle s\mid T^{b^z}(\mu) \mid s^0\rangle =
\sum_{N}{\cal D}_N\langle s\mid T^{N}(\mu) \mid s^0\rangle
\eeq
where ${\cal D}_N$ is the renormalization operator for the
dynamical evolution of the system: it is a projection operator that
samples only the paths of $N$ time steps which have
to be considered in the definition of the effective operator $T^{b^z}$.

The operator  ${\cal D}_N$ is chosen
on the basis of physical considerations: spanning conditions etc.
In addition, ${\cal D}_N$  should satisfy some general
properties in order to preserve the symmetry
or the internal space of the dynamical variables.
For instance, we have to ensure the normalization of the
effective dynamical operator by the property
\beq
\sum_{\{s\}}\sum_{N}
{\cal D}_N\langle s\mid T^{N}(\mu) \mid s^0\rangle = 1
\eeq
Moreover, ${\cal D}_N$
must be consistent with the definition of the renormalization
operator ${\cal R}$: it should describe
dynamical processes among renormalized
variables of the same type of those
given by the operator ${\cal R}$. Finally,
${\cal D}_N$ has to preserve
the form of the dynamical operator $T$ at each scale.
This condition imposes that the time scaling is
consistent with the length scaling used in ${\cal R}$. In this way, it is
possible to map the renormalized system in the old one with renormalized
variables.
The operator ${\cal D}_N$ is therefore defined explicitly as an operator
acting on the paths internal to four site cells. It selects for each $N$
just relaxation paths which consist of $N$ connected non-contemporary
relaxation events that leave the cell without unstable sites. In a
mathematical form it reads as
\beq
{\cal D}_N=\prod_{i\in\{ \alpha'\}}(1-\delta_{2,s_i})
 \prod_{J=0}^{N-1}\sum_{m=1}^{4}\delta(m-\sum_{i\in \{\alpha_J\}}
\delta_{2,s_i})
\eeq
where $\alpha_J$'s are the intermediate cell configurations during
the dynamical evolution and
$\sum_{i\in \{\alpha_J\}}$ denotes the sum over all the sites in the
cells. In the above expression, each delta function acts on a different
intermediate cell eliminating those paths which do not have activity
at each dynamical step. Furthermore, the operator ensures that in the
cell $\alpha'$ ($N$th step) no activity is present; i.e the process has
stopped.
Finally, we have to write the equation that gives the time scaling factor
from the total average over contributing processes to the renormalized
matrix element $\langle 0\mid T^{\prime} \mid 2\rangle$
\beq
g(p_{n})=\frac{1}{\cal N}\sum_{\alpha} W_{\alpha}(\langle\rho\rangle)
\sum_{\alpha'}\sum_N N{\cal D}_N\langle\alpha'\mid T^{N}(p_{n'})
\mid \alpha\rangle
\eeq
where we used the DDRG scheme to explicitly get the stationary weights,
and $\cal N$ is an opportune normalization factor.

The above relations will provide the consistent
rescaling of time by imposing that $b^z=g(p_n^*)$ from which it is possible
to calculate the dynamical critical exponent.

\section{Minimal proliferation calculations and results}

The explicit evaluation of the recursion relations depends on the
choice of the spanning condition. In the following, the scheme used
considers only connected paths that span the cell
from left to right or from top to bottom. This spanning rule implies that only
paths extending over the size of coarse grained length scale contribute
to the renormalized dynamics, and it ensures the connectivity
of the avalanche in the renormalization procedure.

\begin{figure}[h]
\unitlength=1.40mm
\special{em:linewidth 0.4pt}
\linethickness{0.4pt}
\begin{picture}(115.00,55.00)
\put(10.00,30.00){\line(0,1){20.00}}
\put(10.00,50.00){\line(1,0){20.00}}
\put(30.00,50.00){\line(0,-1){20.00}}
\put(30.00,30.00){\line(-1,0){20.00}}
\put(35.00,30.00){\line(0,1){20.00}}
\put(35.00,50.00){\line(1,0){20.00}}
\put(55.00,50.00){\line(0,-1){20.00}}
\put(55.00,30.00){\line(-1,0){20.00}}
\put(60.00,30.00){\line(0,1){20.00}}
\put(60.00,50.00){\line(1,0){20.00}}
\put(80.00,50.00){\line(0,-1){20.00}}
\put(80.00,30.00){\line(-1,0){20.00}}
\put(95.00,30.00){\line(0,1){20.00}}
\put(95.00,50.00){\line(1,0){20.00}}
\put(115.00,50.00){\line(0,-1){20.00}}
\put(115.00,30.00){\line(-1,0){20.00}}
\put(60.00,10.00){\line(0,1){20.00}}
\put(60.00,30.00){\line(1,0){20.00}}
\put(80.00,30.00){\line(0,-1){20.00}}
\put(80.00,10.00){\line(-1,0){20.00}}
\put(95.00,10.00){\line(0,1){20.00}}
\put(95.00,30.00){\line(1,0){20.00}}
\put(115.00,30.00){\line(0,-1){20.00}}
\put(115.00,10.00){\line(-1,0){20.00}}
\put(15.00,45.00){\circle{4.00}}
\put(25.00,45.00){\circle{4.00}}
\put(25.00,35.00){\circle{4.00}}
\put(25.00,35.00){\circle*{2.00}}
\put(15.00,35.00){\circle*{4.00}}
\put(40.00,45.00){\circle{4.00}}
\put(50.00,45.00){\circle{4.00}}
\put(40.00,35.00){\circle{4.00}}
\put(50.00,35.00){\circle{4.00}}
\put(40.00,35.00){\circle*{2.00}}
\put(65.00,35.00){\circle{4.00}}
\put(75.00,35.00){\circle{4.00}}
\put(75.00,45.00){\circle{4.00}}
\put(65.00,45.00){\circle*{4.00}}
\put(75.00,25.00){\circle*{4.00}}
\put(65.00,15.00){\circle{4.00}}
\put(75.00,15.00){\circle{4.00}}
\put(65.00,25.00){\circle*{4.00}}
\put(105.00,20.00){\circle*{4.00}}
\put(105.00,40.00){\circle{4.00}}
\put(105.00,38.00){\vector(0,-1){16.00}}
\put(48.00,35.00){\vector(-1,0){6.00}}
\put(65.00,37.00){\vector(0,1){6.00}}
\put(65.00,33.00){\vector(0,-1){6.00}}
\put(20.00,55.00){\makebox(0,0)[cb]{(a)}}
\put(45.00,55.00){\makebox(0,0)[cb]{(b)}}
\put(70.00,55.00){\makebox(0,0)[cb]{(c)}}
\put(105.00,55.00){\makebox(0,0)[cb]{(d)}}
\put(10.00,28.00){\line(0,-1){8.00}}
\put(30.00,28.00){\line(0,-1){8.00}}
\put(5.00,24.00){\vector(1,0){5.00}}
\put(35.00,24.00){\vector(-1,0){5.00}}
\put(20.00,24.00){\makebox(0,0)[cc]{b}}
\end{picture}
\caption{Example of the renormalization scheme for the
relaxation dynamics. For details see the text.}
\label{fig.1}
\end{figure}
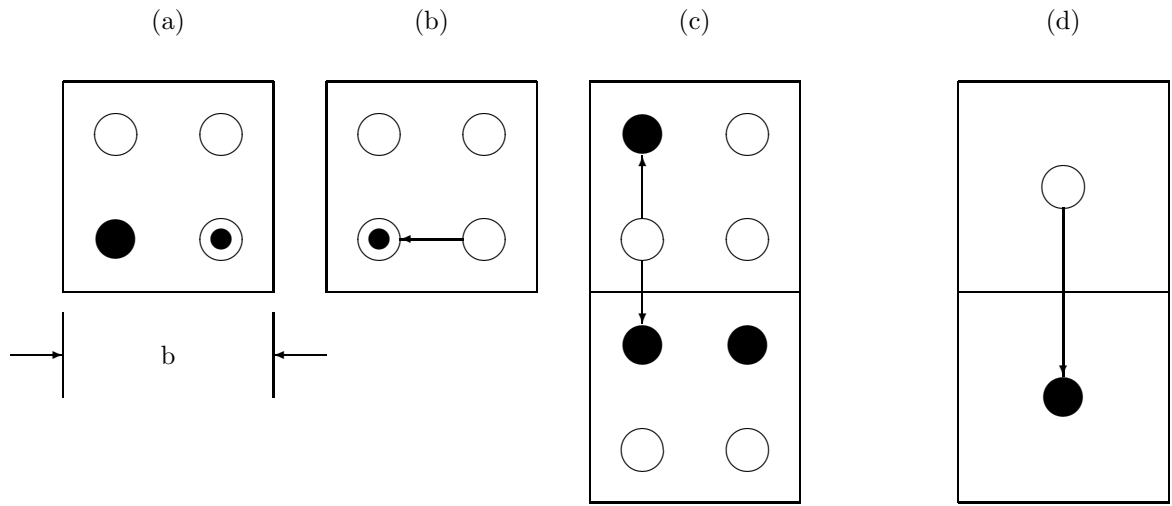

An example of such a path is shown in Fig.(\ref{fig.1}).
In this case $\alpha = 2$
and the path shown refers to the probability that the unstable
sub-cell relaxes towards the other critical sub-cell [Fig.1(b)].
This occurs
with probability $(1/4) p_1$. At this point we consider the probability
that the next relaxation event at the fine scale involves two
neighboring sites,
one inside and the other outside the original cell of
size $b=2$ [Fig.1(c)].This occurs with probability $(2/3) p_2$.
This series of relaxation processes
contributes to the term ${\cal D}_2\langle\alpha'\mid T^{2}(p_{n'})
\mid \alpha=2\rangle_1$  that characterizes
the relaxation processes at the coarse grained scale.
By summing over all the paths that
lead to  a $p'_1$ process one obtains for $\alpha = 2$
\beginwide
\beq
\begin{array}{l}
\sum_{\alpha'}\sum_N {\cal D}_N\langle\alpha'\mid T^{N}(p_{n'})
\mid \alpha=2\rangle_1=  \\(\frac{1}{4}p_{1}+ \frac{1}{6}p_{2})
(\frac{1}{2}p_{1}+ \frac{2}{3}p_{2}\frac{1}{2}p_{3})
+(\frac{1}{6}p_{2}+ \frac{1}{4}p_{3})(\frac{1}{2}p_{1} +
\frac{1}{6}p_{2})+\\
(\frac{1}{6}p_{2}+ \frac{1}{4}p_{3})
(\frac{1}{6}p_{2}+ \frac{1}{4}p_{3})
(\frac{3}{4}p_{1}+ \frac{1}{2}p_{2}\frac{1}{4}p_{3}).
\end{array}
\eeq
\endwide
In a similar way one can also write the expression for the
complete recursion relations. Proliferation effects
due to multiple relaxations
of the same site and sites becoming critical during the
dynamical process are not
considered in this scheme. However, the complete polynomials for
$p'_n$ involve more than two hundred terms, and
we developed a generating function that allows their systematic calculation.

The generating function allows us to find the form of the
renormalized operator $T'$, without writing down the explicit
form of all relaxation paths in the coarse grained cell.
The basic idea of this method is to renormalize the
function which describes all relaxation paths at once,
rather then  the probabilities of separate relaxation paths,
by using the branching structure of  avalanches,

To describe in detail the branching process underlying the large scale
behavior of the sandpile model, we consider the generating
function
\begin{eqnarray}
\lefteqn{\sigma(N,E,S,W)=p_0 + \frac{p_1}{4}(N+E+S+W)}\nonumber\\
& & ~~~~+\frac{p_2}{6}(NE+NS+NW+ES+EW+SW) \label{gf}\\
& & ~~~~+\frac{p_3}{4}(NES+NEW+NSW+ESW)+p_4~NESW, \nonumber
\end{eqnarray}
where symbols $N,E,S$ and $W$ correspond to the north, east, south
and west directions on the square lattice, respectively. The coefficient
in  each term of this polynomial gives the probability for the process to
go in the corresponding directions. The generating function takes into
account all possible relaxation processes in the cell.
It is easy to check directly that this function has the following
properties:
\begin{enumerate}
\item If the argument corresponding to
any direction is replaced by zero, the function counts the relaxation
processes that do not send energy to this direction
(Fig.\ref{fig.2}(b)).
\item If the argument corresponding to any direction
is replaced by unit, the function counts the relaxation
processes whether or not the energy is transfered in that direction
(Fig.\ref{fig.2}(c)).
\item The generating function is normalized so that
$\sigma( 1, 1, 1, 1)=1$.
\end{enumerate}

If there is a critical cell near the relaxing one,
the outgoing energy can initiate the relaxation of the cell.
It is easy to see that we can replace the argument corresponding to
this direction by another generating function corresponding to the
relaxation of the second cell. Finally, we obtain the generating
function of this two-step relaxation process.
For example, the function
$\sigma(N_1,\sigma( N_2,  E_2, S_2, 1 ), S_1, W_1 )$
describes the processes where the cell $1$ relaxes first.
Then, if the energy goes to the direction $N_1$ it initiates
the relaxation of the cell $2$ (fig.\ref{fig.2}(d)).
The symbols $N_i, E_i, S_i, W_i$ denote the directions
outgoing from the cell $i$.

Using these properties we can write down
the generating function $\Sigma_\alpha$,
counting the relaxation processes in the block
that consists of four cells for different $\alpha$.
To this end, we must take into account
only the processes which match the spanning
condition. Therefore, it is necessary to eliminate all
processes, in which only one cell relaxes.
As the coefficients of the polynomial $\Sigma_\alpha$
have the meaning of probabilities, they should finally be normalized
by the condition
\beq
\Sigma_\alpha(1,...,1)=1.
\eeq

The generating function corresponding to the relaxation processes inside
the block with $\alpha=2$ is
\begin{eqnarray}
& &~{\Sigma_2=} \nonumber \\
& &~ \{\sigma(\sigma(N_2,1,1,W_2),1,S_1,W_1)-\sigma(0,1,S_1,W_1)\}+\nonumber \\
& &~ \{\sigma(N_2,1,\sigma(1,1,S_1,W_1),W_2)-\sigma(N_2,1,0,W_2)\}+\\
& &~c.p.)/Z_2,\nonumber
\end{eqnarray}
where $Z_2$ is a normalization factor chosen so
that $\Sigma_2(1,...,1)=1$. To write this function we start from the left
down cell and define the arguments of the $\sigma$-function corresponding
to the toppling of this cell. For the process definitely spans the block,
the left up cell should topple and we write another $\sigma$-function
instead of the symbol $N_1$. By going eastward the process will terminate
inside the block and this branch of the toppling process cannot affect
the neighboring blocks. Hence, we should write the number $1$ instead of
the symbol $E_1$. The other symbols $S_1$ and $W_1$
correspond to the branches
of the relaxation process that goes immediately out
of the initial block of cells.
To consider only the processes that span the block, we must subtract the
$\sigma$-function describing processes that do not send the energy
from the first critical cell to the second one.
Then, we add analogous  $\sigma$-functions
for the processes starting from the relaxation of the left upper cell.
The term $c.p.$ denotes all possible cyclic permutations of
the critical cells inside the block. Analogously, we can write
$\sigma$-functions of all other types of blocks.

To obtain complete generating function $\Sigma$ for the block
of four cells, we should sum up all $\Sigma_\alpha$-functions with the
weights of blocks and normalize the result
\beq
\Sigma =\frac{1}{Z} \sum_\alpha W_\alpha \Sigma_\alpha.
\eeq
Now, to transform the $\Sigma$-function from the block of four cells
at the scale $b^k$ to larger cell at the next scale $b^{k+1}$,
we replace the directions $N_1,N_2,...$ outgoing from the initial block
by the new arguments corresponding to the directions $N,...$
outgoing from the new renormalized cell.
In other words, two bonds that connect the neighboring blocks are
coupled to the only bond on the lattice at the next scale, as is shown
in fig.\ref{fig.3}.
Eventually, we obtain the following generating function:
\begin{eqnarray}
& &~{\Sigma(N,E,S,W)=\frac{W_2 \Sigma_2  +  W_3 \Sigma_3 + W_4 \Sigma_4}Z}\\
& &~ \nonumber\\
& &~{\rm where}\nonumber\\
& &~{\Sigma_2=} \nonumber \\
& &~ (\sigma(\sigma(N,1,1,W),1,S,W)-\sigma(0,1,S,W))+\nonumber\\
& &~ (\sigma(N,1,\sigma(1,1,S,W),W)-\sigma(N,1,0,W))+c.p.)/Z_2,\nonumber\\
& &~ \\
& &~{\Sigma_3(N,E,S,W)=} \nonumber \\
& &~ (\sigma(\sigma(N,\sigma(N,E,1,1),1,W),1,S,W)-\sigma(0,1,S,W)\nonumber \\
& &~ +\sigma(\sigma(N,1,1,W),\sigma(1,E,S,1),S,W)-\sigma(0,0,S,W)\nonumber  \\
& &~ +\sigma(1,\sigma(\sigma(N,E,1,1),E,S,1),S,W)-\sigma(1,0,S,W) \nonumber\\
& &~ +c.p.)/Z_3 \nonumber\\
& &~ \nonumber \\
& &~\Sigma_4(N,E,S,W)= \nonumber\\
& &~(\sigma(\sigma(N,\sigma(N,E,\sigma(1,E,S,1),1),1,W),0,S,W)+\nonumber \\
& &~\sigma(0,\sigma(\sigma(N,E,1,\sigma(N,1,1,W)),E,S,1),S,W)-  \nonumber \\
& &~2\sigma(0,0,S,W) + (\sigma(1,1,S,W)-\sigma(1,0,S,W)- \nonumber \\
& &~\sigma(0,1,S,W)+\sigma(0,0,S,W))                     \nonumber \\
& &~(\sigma(N,0,1,W)\sigma(0,E,S,1)+\sigma(N,E,1,1)      \nonumber \\
& &~(\sigma(N,0,1,W)(\sigma(1,E,S,1)-\sigma(0,E,S,1))+   \nonumber \\
& &~\sigma(0,E,S,1)(\sigma(N,1,1,W)-\sigma(N,0,1,W))+    \nonumber \\
& &~(\sigma(1,E,S,1)-\sigma(0,E,S,1))                    \nonumber \\
& &~(\sigma(N,1,1,W)-\sigma(N,0,1,W))))+c.p.)/Z_4        \nonumber
\end{eqnarray}
Here, $Z_i$ and $Z$ are the normalization factors and $c.p.$ denote the expressions
obtained from the previous polynomial by all possible cyclic
permutations of its arguments. This generating function is the polynomial
that contains only the first and second powers of its arguments.
The last terms correspond to the processes when two energy portions are
transferred from the initial block to the neighboring block by the two
paths. However, according to the RG strategy, these processes should be
considered as the transfer of the coarse grained energy portion
at the larger scale.
Therefore, all second powers of the arguments should be replaced
by the first ones.
The result obtained is the generating function describing the relaxation
of renormalized cells. It depends on the same products of its arguments
as the generating function for initial cells, but the coefficients of
this polynomial are different and equal to the probabilities of relaxation
processes that send energy to the given direction at the new scale.
Taking these coefficients, we obtain the sought recursion
relations which link the parameters of the cell
at the scale $b^k$ with the same parameters
at the scale $b^{k+1}$
\beq
{\vec P}{(k+1)}=\vec f({\vec P}{(k)},\rho{(k)})
\label{rec}
\eeq
The above set of RG equations supplemented with the
driving condition
\beq
\langle\rho^{k+1}\rangle=\frac{1}{\sum_n np^{k+1}_n}
\eeq
define completely the DDRG recursion relations for sandpile
models.
Given this scheme the flow diagram and the relative fixed point in the
parameters space $(\rho,{\vec P})$ can be studied. We consider here
the calculation scheme implemented with ${\vec
P}=(p_0,p_1,p_2,p_3,p_4)$. Despite the enlargement of the phase space by
including the proliferation characterized by the probability $p_0$ the flow in 
the phase space is very similar to those obtained in ref.\cite{pvz}, where 
this parameter was not considered. A single attractive fixed point
is obtained and the  numerical value of this
fixed  point is very close to that obtained in the
approach of Ref.~\cite{pvz}. The complete
attractiveness of the fixed point corresponds to the lack of relevant
scaling field, i.e. control parameter. This must be the case, because we
implement our RG scheme in the infinite time scale separation limit.
This implies the relevant scaling field are constrained to their
critical values. In other words we are restricting the study of the
system on its critical surface.
In Tab.~I we report the results obtained within this calculation scheme.
The single fixed point is the signature of a single universality for all
the non-directed sandpile models. This is questioned from numerical
simulations which show evidences this could  not be the case \cite{ben}.
It appears, in fact, that BTW model and Manna model belong to different
universality classes. This distinction does not appear in the present
proliferation scheme. 
The present approach, however, can not provide a definitive 
settling of this issue. 
Mainly,  it depends on the fact that we  still
are neglecting some proliferations such as 
the possibility of multiple topplings, that could be relevant for the 
identification of different universality classes. 

The avalanche exponent $\tau$ can be obtained directly from the fixed point
parameters. By using the discrete length scale $b^{(k)}=2^k$ and the
avalanche distribution in the form $P(r)dr\approx r^{(1-2\tau)}dr$ we can
define the probability that the relaxation process spans the cell of size
$b^{(k)}$ and dies at the neighboring cells not extending over the
scale $b^{(k+1)}$
\begin{equation}
K=\int_{b^{(k-1)}}^{b^{(k)}}P(r)dr/
\int_{b^{(k-1)}}^{\infty}P(r)dr=1-2^{2(1-\tau)}
\label{K11}
\end{equation}
Asymptotically $(k\rightarrow\infty)$ we can express $K$ in
terms of fixed point parameters $\rho^*$ and $p_i^*$ in the following way:
\begin{eqnarray}
\lefteqn{K= p_0^* + p_1^*(1-\rho^*)+p_2^*(1-\rho*)^2}\nonumber\\
& &~~~+p_3^*(1-\rho^*)^3+p_4^*(1-\rho^*)^4.
\label{K12}
\end{eqnarray}
This equation gives the total probability that a relaxation process
occurs without triggering other sites, and therefore it does not extend
on length scales larger than that of a single cell.
Using these two expressions, Eqs.\ (\ref{K11},\ref{K12}), the exponent $\tau$ is
given by the formula
\begin{equation}
\tau=1-\frac{1}{2}\frac{\ln(1-K)}{\ln2}=1.262
\end{equation}
in excellent agreement with the proposed value $\tau=5/4$\cite{PKI} and
large scale numerical simulations\cite{manna,grasma,lubeck}.
The obtained value is also in good agreement with the value obtained in
the calculation of ref~\cite{pvz}, showing the robustness of the method
with respect to different proliferation scheme.
In order to overcome some of the approximations considered so far, we
will present in the next section an improved scheme which takes into
account a wider set of dynamical parameters. This scheme allows us to
study also the critical behavior at the boundary of the system.

\section{Extended kinetic equation scheme}

In this section we treat more explicitly the dynamics of
the original sandpile model.
Considering the evolution of stable cells,
we can take into account some of  the processes which were neglected
in the scheme discussed in the previous section.

To keep the connection
with the original formulation of the sandpile model, we will characterize
the static properties of a cell by four quantities
\begin{equation}
{\vec N}{(k)} = (n_{\rm A}, n_{\rm B}, n_{\rm C}, n_{\rm D}),~~~
n_{\rm A} + n_{\rm B} + n_{\rm C} + n_{\rm D} = 1,
\end{equation}
which are nothing but the probabilities for a cell to behave
like a site on the initial lattice with a height 1, 2, 3 or 4,
respectively, in the coarse grained dynamics, {\it i.e.} the addition of a "coarse grained
particle" to the cell transforms it to the next one in the alphabet.
For example, the cell B characterized by the vector (0,1,0,0) will
be transformed to the cell C with the vector (0,0,1,0).
The last variable $n_{\rm D}$ is the probability for the cell to behave
like a critical one in a sense that the addition of a "coarse grained particle"
to the cell induces relaxations into some neighboring cells
or, in other words, subrelaxation processes on a
minimal scale span the cell and transfer energy to some of its neighbors.

As we stressed in the previous sections,
independently of the dynamics of the model at the minimal scale,
each critical cell is characterized by the vector
\beq
{\vec P}(k) = (p_1, p_2, p_3, p_4),~~~~~  p_1 + p_2 + p_3 + p_4 = 1,
\label{p}
\eeq
that gives the probabilities for the energy to go to $1,2,3$ or $4$
neighboring cells after the relaxation of the critical cell.
Here, because we have already enlarged the phase space by introducing
the densities $n_\alpha$, we do not include $p_0$ in the calculation scheme.

In this framework the coarse grained dynamics of the sandpile model can be
represented as the following branching process on the sublattice
${\cal L}_b$:
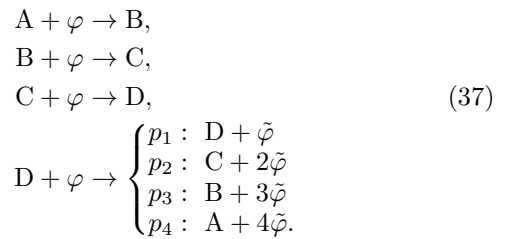
\begin{eqnarray}
{\rm A} + \varphi & \rightarrow & {\rm B}, \nonumber\\
{\rm B} + \varphi & \rightarrow & {\rm C}, \nonumber\\
{\rm C} + \varphi & \rightarrow & {\rm D}, \\
{\rm D} + \varphi & \rightarrow &
\left\{
\begin{minipage}{2.3cm}
$p_1:~ {\rm D} +  \tilde{\varphi}$\\
$p_2:~ {\rm C} + 2\tilde{\varphi}$\\
$p_3:~ {\rm B} + 3\tilde{\varphi}$\\
$p_4:~ {\rm A} + 4\tilde{\varphi}.$
\end{minipage}
\right.  \nonumber
\end{eqnarray}
Here, $\varphi$ and $\tilde{\varphi}$ denote the "coarse grained particles"
obtained by the cell and the particles transferred to the neighboring cells,
respectively.

These processes can be formally reinterpreted as an irreversible chemical
reaction which takes place at each cell of the sublattice ${\cal L}_b$.
Now the coarse grained variables $n_{\rm A}, n_{\rm B}, n_{\rm C}, n_{\rm D}$ and
$n_\varphi$ denote the concentrations of the respective species
A, B, C, D, and $\varphi$. Following to standard prescriptions of the
chemical physics we can write down kinetic equations corresponding to this
scheme of chemical reactions
\begin{eqnarray}
\dot{n}_{\rm A} &=& n_\varphi~ (p_4~ n_{\rm D} - n_{\rm A}), \label{chem1}\\
\dot{n}_{\rm B} &=& n_\varphi~ (p_3~ n_{\rm D} + n_{\rm A} - n_{\rm B}), \label{chem2}\\
\dot{n}_{\rm C} &=& n_\varphi~ (p_2~ n_{\rm D} + n_{\rm B} - n_{\rm C}), \label{chem3}\\
\dot{n}_{\rm D} &=& n_\varphi~ (p_1~ n_{\rm D} + n_{\rm C} - n_{\rm D}), \label{chem4}\\
\dot{n}_\varphi &=& n_\varphi~ (\bar{p}~n_{\rm D} - 1) + \bar{p}~ \nu \nabla^2 (n_\varphi n_{\rm D}) +
\eta ({\bf r}, t) \label{chem5}
\end{eqnarray}
where $\bar{p}=p_1+2p_2+3p_3+4p_4$ is equal to the average number of
particles leaving the cell on toppling and ${\bf r}$ is the position vector
of the cell in the 2D space. The noise term $\eta({\bf r}, t)$,
being non-negative, mimics the random addition of particles to the system.
The diffusion term $\nabla^2 (n_\varphi n_{\rm D})$ describes the transfer
of particles into the neighboring cells, and the diffusion coefficient $\nu$
for the discrete Laplacian on the square lattice is equal to $1/4$.

The only mobile specie in this scheme of reactions is $\varphi$ and it
is the field $n_\varphi$ which describes the dynamics of avalanches. When it
is equal to zero, all toppling processes die. Then, due to the noise
term $\eta({\bf r},t)$, particles are added randomly into the system
initiating a branching process directed to the open boundary of the
system. This process mutates species in the cells it has visited and topples
the critical ones. Finally, the system will reach the steady state where
the probability that the activity will die is on average balanced by the
probability that the activity will branch. Thus, the chain reaction maintains
this stationary state and all further avalanches cannot change the
concentrations of species A, B, C, and D. Therefore, the steady
state is characterized by the conditions that
\begin{equation}
\dot{n}_{\rm A}=\dot{n}_{\rm B}=\dot{n}_{\rm C}=\dot{n}_{\rm D}=0
\end{equation}
and Eqs.\ (\ref{chem1}-\ref{chem4}) lead to the following relationships
between concentrations of species ${\vec N}{(k)}$ at the stationary state and
branching probabilities ${\vec P}{(k)}$
%\begin{mathletters}
\begin{eqnarray}
n_{\rm A}^* &=& p_4/\bar{p},  \label{balanceA}\\
n_{\rm B}^* &=& (p_3+p_4)/\bar{p},  \label{balanceB}\\
n_{\rm C}^* &=& (p_2+p_3+p_4)/\bar{p}, \label{balanceC}\\
n_{\rm D}^* &=& (p_1+p_2+p_3+p_4)/\bar{p}=1/\bar{p}~. \label{balanceD}
\end{eqnarray}
%\end{mathletters}
The relation (\ref{balanceD}) between the probability $n_{\rm D}^*$
and branching probabilities ${\vec P}{(k)}$
can also be derived from the assumption that at the stationary state the
flow of particles in a cell was on average balanced by the flow of
particles out of the cell.

Thus, we have found the driving conditions for the sandpile models.
Using them we can link the statistic weights of
any static configurations of a cell with the dynamic parameters.
Now, we can realize the renormalization procedure
described previously. To this end, we must consider
all types of blocks of four cells, whose relaxation
matches the spanning condition. Such blocks
and some of their relaxation schemes
are shown in fig.\ref{fig.4}.
While the previous scheme deals only with the cells
being critical before the relaxation of the block,
this one allows us to consider the cells becoming
critical during the relaxation(fig.\ref{fig.4}(c)).

To obtain the recursion relations by the method
presented in the previous section, it is necessary
to calculate the statistic weights of all configurations
considered. The statistic weight of the block
is given by the product of probabilities
$n_i$ for all cells in the block, multiplied by
the number of different blocks with the same relaxation
schemes. Thus, the following weights must be ascribed to the
blocks shown in fig.\ref{fig.4}:
\begin{eqnarray}
W_a&=&4 n_{\rm D}^2 (n_{\rm A}+n_{\rm B}+n_{\rm C})^2,\label{weight1}\\
W_b&=&4 n_{\rm D}^3 (n_{\rm A}+n_{\rm B}), \label{weight2}\\
W_c&=&4 n_{\rm D}^3 n_{\rm C}, \label{weight3}\\
W_d&=&  n_{\rm D}^4.\label{weight4}
\end{eqnarray}
Expressing them through the probabilities $\vec P$,
by using driving conditions (\ref{balanceA}-\ref{balanceD}),
we obtain the complete system of renormalization equations.
\beq
\vec P{(k+1)}=\vec f(\vec P{(k)})
\label{bulk}
\eeq
Given this set of RG transformations we can study how the system evolves
under successive doubling of length scale. The final result is independent
upon the initial conditions $n_i$ and $p_i$ at the minimal scale.
Also in this case the fixed point is attractive in the whole phase space and
the system evolves spontaneously toward the fixed point values $p_i^*$ and
$n_i^*$ shown in Tables \ref{fixed_point} and \ref{height_prob}. These
results can be compared with the exact ones obtained for the
sandpile\cite{P}. The exact height probabilities for the
sandpile are reported in Tab.\ref{height_prob} and compared with
our RG results.
In order to calculate the avalanche exponent we can use the eq.(\ref{K11})
by expressing $K$ in terms of the fixed point parameters in the
following way:
\begin{eqnarray}
\lefteqn{K=p_1^*(1-n_{\rm D}^*)+p_2^*(1-n_{\rm D}^*)^2}\nonumber\\
& &~~~+p_3^*(1-n_{\rm D}^*)^3+p_4^*(1-n_{\rm D}^*)^4.
\label{K2}
\end{eqnarray}
More generally the probability (\ref{K2}) should be represented via
$\sigma$-function
\beq
K=\sigma^*(1-n_D^*(N),1-n_D^*(E),1-n_D^*(S),1-n_D^*(W))
\eeq
where $n_D(N),n_D(E),n_D(S)$ and $n_D(W)$ are the concentrations of
critical cells at the nearest neighbors.
By using the fixed point values we finally obtain $\tau=1.248$, that
again is in very good agreement with the proposed value $5/4$ \cite{PKI} and
numerical simulations\cite{manna,grasma,lubeck}.
Also in this case the scheme results in a single universality class for the
sandpile models. In the extended scheme we include part of the proliferations
by allowing different heights, and taking into account that some of the sites
becomes critical during the relaxation event. We do not, however, consider
multiple relaxation of the same sites during the spanning time, nor we allows
a renormalization of the energy transfer ($\delta E$). This parameters could 
be important in the case of the Manna model as pointed out in
ref.\cite{ben}. Work is in progress to extend the present DDRG scheme in
order to include also these further proliferations.

\section{The boundary critical Properties}

Since the critical properties of the sandpile model are quite similar to those
of second order phase transitions, we
proceed here along the same lines followed in the study of equilibrium
critical phenomena. In particular  we determine also the
surface critical exponents, which in general differ from the
bulk ones. This
is of special importance in the two-dimensional case where conformal
field theory connects surface and bulk properties of the model
\cite{cardy-84}.

\subsection{Open boundary}
The fact that the boundary is open means that after the relaxation
of the boundary site the energy can leave the system.
We consider the critical energy of sites at the open boundary
to be $ E_c = 4$.
It is more convenient to consider
the boundary lying along the diagonal of the lattice
and construct the renormalized cell in the following way.
We consider the block of cells $2 \times 2$
that contains one bulk, two boundary and one external
cells as shown in fig. \ref{fig.5}.
For the critical properties of the model at large scales
does not depend on the local structure of the lattice, the results
obtained should not depend on the specific choice of the boundary.
To describe the boundary cells at an arbitrary scale, we introduce
the vectors
\bea
{\vec N}^o(k) = (n^o_{\rm A}, n^o_{\rm B}, n^o_{\rm C}, n^o_{\rm D}),\nonumber\\
n^o_{\rm A} + n^o_{\rm B} + n^o_{\rm C} + n^o_{\rm D} = 1,
\eea
\bea
{\vec P}^o(k) = (p^o_1, p^o_2, p^o_3, p^o_4),\nonumber\\
p^o_1 + p^o_2 + p^o_3 + p^o_4 = 1,
\label{popen}
\eea
which have the same meaning as in the previous section.
The kinetic equations
and feedback relations coincide with the bulk ones.
Thus, it is only necessary to find the correct form of
recursion relations.
To write the generating function for the boundary block,
we can use again the generating functions describing the relaxations
of bulk and boundary cells. Also we have to introduce a special
generating function corresponding to the relaxation of the unphysical
external cell. Since there are not processes transferring energy from
the external half-plane of the lattice to the internal one, we require
that the external cell immediately transfers the energy outside the lattice.
Thus, we provide conservation of the flaw of energy
through the boundary in the scale transformation.
For the block shown in fig.\ref{fig.5}
the generating function of relaxation of the external
cell has the simple form
\beq
\sigma^{out}=\frac{N + E}2.
\eeq
Statistical weights of the static boundary configurations are given by the
product of the probabilities of one bulk and two boundary cells.
Now, the coefficients of the generating function of the renormalized cell
gives us the recursion relations
\beq
\vec P^o{(k+1)}=\vec f_o(\vec P^o{(k)},\vec P{(k)}).
\label{openrec}
\eeq
Together with the bulk recursion relations of Eq.~(\ref{bulk}),
Eq.(\ref{openrec})
represents the complete system of the
renormalization equations for
the case of the open boundary.
This system also has only one fixed point.
The obtained fixed point parameters are
given in  Tables~\ref{fixed_point} and \ref{height_prob}.
The comparison of fixed point height probabilities
with the exact values obtained for ASM \cite{I}
shows rather good agreement.

To calculate the boundary critical
exponent we should use the  generating function describing the relaxation
of the boundary cell in the fixed-point of RG flow.
It is  given by eq.~(\ref{K11}), where
\beq
K=\sigma^*_o(1,1,1-n_{\rm D}^o,1-n_{\rm D}^o).
\eeq
This indicates that the toppling of coarse grained boundary cell
should stop in two neighboring internal cells.
The result $\tau_o=1.486$ is very close to $\tau_o=\frac{3}{2}$ calculated
exactly in \cite{IKP} for open boundary of ASM.
Such a good agreement is probably to be ascribed to the fact
that in the boundary avalanche in ASM
each site topples only once \cite{wave}.
Therefore renormalization scheme that does not take into account
multiple topplings gives most realistic results
near open boundary.

\subsection{Closed boundary}

Let us consider the sandpile model on the half-plane.
The edge of this half-plain is the closed boundary
directed along one of the lattice axes.
Each boundary site has three neighboring sites. Two
of them also belong to the boundary.
The fact that the boundary is closed means that after the relaxation
of a boundary site energy does not leave the lattice,
being distributed among the neighboring sites.
Since the energy can leave the boundary site only
in three directions, it is quite natural to take
critical energy of a boundary site ${E}_c=3$.

In order to follow the RG strategy,
we again perform the site-to-cell
transformation, replacing the block of
four cells by a single cell
at the larger scale.
The cells at an arbitrary scale can be considered as either
boundary or bulk ones.
While the former consist only of bulk cells at the
smaller scale, the latter include both the bulk and the
boundary cells of smaller size.
The renormalization of the bulk cell is described by the system of
RG equations obtained in the previous section.
Let us introduce the description
for the dynamics of boundary cells at an arbitrary scale.
The static states of a boundary cell can be represented by three symbols
$\rm A,B,C$, which correspond to energy values $E=1,2,3$ of boundary
sites on the initial lattice. The addition of energy
transforms the cell from the state $\rm A$
into the state $\rm B$ and the cell from the state $\rm B$ into the state $\rm C$.
The cell in the state $\rm C$ is critical.
The addition of energy initiates its relaxation
when the cell turns into the states A or B or remains
in the state C, sending the energy to three, two or one neighboring
cells, respectively.
The probabilities for the cell on the closed boundary
to be in one of the three states is given by the vector
\beq
{\vec N}^{c}{(k)} = (n^{c}_{\rm A}, n^{c}_{\rm B}, n^{c}_{\rm C} ),~~~~
n^{c}_{\rm A} + n^{c}_{\rm B} + n^{c}_{\rm C} = 1.
\eeq
In the same way, the probabilities for the energy to be transferred in
one, two or three directions is given by the vector
\beq
{\vec P}^{c}{(k)} = (p^{c}_1, p^{c}_2, p^{c}_3),~~~~~
p^{c}_1 + p^{c}_2 + p^{c}_3 = 1.
\label{pálose}
\eeq
The relaxation process at the boundary cell
can be represented as follows:
\bea
{\rm A} + \varphi & \rightarrow & {\rm B}, \nonumber\\
{\rm B} + \varphi & \rightarrow & {\rm C}, \nonumber\\
{\rm C} + \varphi & \rightarrow &
\left\{
\begin{minipage}{5cm}
$p^c_1:~ {\rm C} + 1\tilde{\varphi}$\\
$p^c_2:~ {\rm B} + 2\tilde{\varphi}$\\
$p^c_3:~ {\rm A} + 3\tilde{\varphi}.$
\end{minipage}
\right.  \nonumber
\eea
where $\varphi$ and $\tilde{\varphi}$ are the energy obtained by the cell
and transferred to the neighboring cell, respectively.
Hence, we can write the following kinetic equations for the energy
transfer:
\bea
\dot{n^c}_{\rm A} &=& n_\varphi~ (p^c_3~ n^c_{\rm C} - n^c_{\rm A}), \label{chem1á}\\
\dot{n^c}_{\rm B} &=& n_\varphi~ (p^c_2~ n^c_{\rm C} + n^c_{\rm A} - n^c_{\rm B}), \label{chem2á}\\
\dot{n^c}_{\rm C} &=& n_\varphi~ (p^c_1~ n^c_{\rm C} + n^c_{\rm B} - n^c_{\rm C}), \label{chem3á}\\
\dot{n^c}_\varphi &=& n_\varphi~ (\bar{p}c~n_{\rm D} - 1) + \bar{p}c~ \nu \nabla^2 (n_\varphi n^c_{\rm C}) +
\eta ({\bf r}, t) \label{chem5á}
\eea
Here, the discrete Laplacian $\Delta$ must be understood with the
Neumann boundary conditions.
The steady state corresponds to the conditions,
$\dot{{\vec N}^c} = 0 $.
This leads us to the following driving conditions for the closed
boundary:
\bea
n^c_{\rm A} &=& {p^c}_3/\bar{p},  \label{cbalanceA}\\
n^c_{\rm B} &=& ({p^c}_2+p^c_3)/\bar{p^c},  \label{cbalanceB}\\
n^c_{\rm C} &=& ({p^c}_1+p^c_2+p^c_3)/\bar{p^c}=1/\bar{p^c}~. \label{cbalanceC}\\
\eea

To perform the standard renormalization procedure described above
and find the recursion relations, the generating function method can be
employed. To this end, we introduce the generating function
for the relaxation of a cell on the closed boundary as the following
polynomial:
\bea
\sigma_c( N, E, W )=\frac{p^c_1}{3}( N + E + W ) +
~~~~~~~~~~~~~\nonumber\\
\frac{p^c_2}{3}( NE + NW + EW ) + p^c_3 NEW.
\label{gfc}
\eea
The general idea of the generating function for the block
of four cells is the same as in the bulk case. The difference is that
the generating functions for boundary blocks
oriented differently with respect to the boundary should be
calculated separately and cannot be obtained by simple cyclic permutations
of the arguments.
Finally, applying the renormalization procedure with the use of the feedback
relations (\ref{cbalanceA}-\ref{cbalanceC}), we obtain the recursion relations
\beq
\vec P^c{(k+1)}=\vec f_c(\vec P^c{(k)},\vec P{(k)}),
\label{closerec}
\eeq
where $\vec P{(k)}$ matches the bulk recursion relations of Eq.(\ref{bulk}).
The obtained height probabilities (Table \ref{height_prob})
are in good agreement with those calculated exactly in
the case of ASM with closed boundary.
To calculate the critical exponent $\tau_c$ describing
the distribution of avalanches
near a closed boundary, we use the Eq.(\ref{K11}), expressing $K$
through the fixed point generating function of the boundary cell:
\beq
K=\sigma^*_c(1-n^c_{\rm C},1-n_{\rm D},1-n^c_{\rm C}).
\eeq
The critical exponent results to be $\tau_c=1.239$.
The correction to the bulk critical exponent
due to the half-plane geometry was
not presented before.
The agreement of the bulk and open boundary exponents
with exact values allows us to expect such an accuracy
in the case of the closed boundary.

\section{Discussion and Conclusions}

In this paper we have presented the detailed application
of the DDRG to the sandpile model. We have concentrated on
the BTW model which we have studied using schemes of increasing
complexity. In the simple scheme the sites are subdivided
in three states (stable, critical and active)
and the RG transformation acts on the energy transfer probabilities
$p_i$ \cite{pvz}. The scheme is then extended in order to treat
explicitly the four states probability densities $n_\alpha$, which can
be obtained self-consistently. The fixed point values of $ n_\alpha$
and $p_i$ are in good agreement with exact results. In addition,
we compute the critical exponent $\tau$ describing the avalanche
size distribution. The result is in good agreement with numerical
and analytical estimations and appears to be robust with respect
to the different approximations. Finally, we study the boundary
scaling of the sandpile model, obtaining results in good agreement
with exact results.
The present analysis shows that the DDRG is a complete and
systematic tool to study nonequilibrium systems in general
and sandpile models in particular.

\section*{Acknowledgments}
E.V.I. and A.M.P are grateful to V.B.~Priezzhev for fruitful discussions.
A.V. is indebted to J.M.J. van Leeuwen for very interesting discussions.
A.V. and S.Z. thank V. Loreto and L. Pietronero with whom they have
collaborated on part of the work described here.
A.M.P.  was partially supported by the Russian Foundation for
Basic Research under grant No. 97-01-01030
and by INTAS grant No. 96-690.
A.M.P. gratefully acknowledges support from
the International Soros Science Educational Program.
The Center for Polymer Studies is supported by NSF.

\newpage

\begin{narrowtext}
\begin{table}[t]
\begin{tabular} {lccccc}
   & $p^*_0$ & $p_1^*$ & $p_2^*$ & $p_3^*$ & $p_4^*$ \\ \hline

$\rho^*=$0.595 & 0.091 & 0.345 & 0.379 & 0.161 & 0.024 \\
\end{tabular}
\caption{The fixed point probabilities for the energy transfer
from the relaxing cell, including the probability $p_0$.}
\label{pzero}
\end{table}

\begin{table}[t]
\begin{tabular} {lcccc}
   & $p_1^*$ & $p_2^*$ & $p_3^*$ & $p_4^*$ \\ \hline

Bulk& 0.295 & 0.435 & 0.229 & 0.0414 \\ \hline

Open  &0.142&0.417&0.351&0.0899 \\ \hline

Closed  &0.526&0.394&0.0799& \\
\end{tabular}
\caption{Relaxation probabilities in the extended RG scheme.}
\label{fixed_point}
\end{table}

\begin{table}[t]
\begin{tabular} {llcccc}
&& $n_{\rm A}$ & $n_{\rm B}$ & $n_{\rm C}$ & $n_{\rm D}$ \\
&&&&& \\
\hline

Bulk        &RG&0.0205&0.134 &0.349&0.496 \\  \cline{2-6}
&Exact\cite{P}& 0.0736 & 0.174 & 0.306 & 0.446 \\ \hline

Open    &RG&0.0377&0.184&0.359&0.419 \\  \cline{2-6}
boundary      &Exact\cite{I}&0.104&0.217&0.316&0.363 \\ \hline

Closed   &RG&0.0514&0.305&0.643& \\  \cline{2-6}
boundary      &Exact\cite{I}&0.113&0.318&0.568&\\
\end{tabular}
\caption{Height probabilities in the stationary state.}
\label{height_prob}
\end{table}
%\endwide
\end{narrowtext}

\begin{table}[t]
\begin{tabular} {lccc}
   & $\tau$ & $\tau_o$ & $\tau_c$ \\ \hline

RG& 1.248 & 1.486 & 1.239  \\ \hline

Exact  &1.25\cite{PKI} &1.5\cite{IKP} &?  \\
\end{tabular}
\caption{Critical exponents $\tau$,$\tau_o$ and $\tau_c$
for the  avalanche size distribution in the
bulk, open and closed boundaries, respectively.}
\label{exponents}
\end{table}

\end{multicols}

\widetext
%%%%%%%%%%%%%%%%%%%%%%%%%%%%%%%%%%%%%%%%%%%%%%%%%%%%%%%%%%%%%%%
\newpage

%%%%%%%%%%%%%%%%%%%%%%%%%%%%%%%%%%%%%%%%%%%%%%%%%%%%%%%%%%%%%%%%
\newpage
\begin{figure}[h]
\unitlength=1.00mm
\special{em:linewidth 0.4pt}
\linethickness{0.4pt}
\begin{picture}(130.00,152.00)
\put(33.00,45.00){\circle{4.00}}
\put(53.00,45.00){\circle{4.00}}
\put(35.00,45.00){\vector(1,0){16.00}}
\put(33.00,47.00){\vector(0,1){8.00}}
\put(33.00,43.00){\vector(0,-1){8.00}}
\put(31.00,45.00){\vector(-1,0){8.00}}
\put(53.00,47.00){\vector(0,1){8.00}}
\put(53.00,43.00){\vector(0,-1){8.00}}
\put(55.00,45.00){\vector(1,0){8.00}}
\put(33.00,61.00){\makebox(0,0)[cc]{$N_1$}}
\put(18.00,45.00){\makebox(0,0)[cc]{$W_1$}}
\put(33.00,30.00){\makebox(0,0)[cc]{$S_1$}}
\put(53.00,61.00){\makebox(0,0)[cc]{$N_2$}}
\put(53.00,30.00){\makebox(0,0)[cc]{$S_2$}}
\put(68.00,45.00){\makebox(0,0)[cc]{$E_2$}}
\put(35.00,130.00){\circle{4.00}}
\put(35.00,132.00){\vector(0,1){8.00}}
\put(37.00,130.00){\vector(1,0){8.00}}
\put(35.00,128.00){\vector(0,-1){8.00}}
\put(33.00,130.00){\vector(-1,0){8.00}}
\put(35.00,145.00){\makebox(0,0)[cc]{$N$}}
\put(20.00,130.00){\makebox(0,0)[cc]{$W$}}
\put(35.00,115.00){\makebox(0,0)[cc]{$S$}}
\put(50.00,130.00){\makebox(0,0)[cc]{$E$}}
\put(115.00,130.00){\circle{4.00}}
\put(117.00,130.00){\vector(1,0){8.00}}
\put(115.00,128.00){\vector(0,-1){8.00}}
\put(113.00,130.00){\vector(-1,0){8.00}}
\put(100.00,130.00){\makebox(0,0)[cc]{$W$}}
\put(115.00,115.00){\makebox(0,0)[cc]{$S$}}
\put(130.00,130.00){\makebox(0,0)[cc]{$E$}}
\put(76.00,130.00){\circle{4.00}}
\put(76.00,132.00){\vector(0,1){8.00}}
\put(78.00,130.00){\vector(1,0){8.00}}
\put(76.00,128.00){\vector(0,-1){8.00}}
\put(74.00,130.00){\vector(-1,0){8.00}}
\put(76.00,145.00){\makebox(0,0)[cc]{$N$}}
\put(61.00,130.00){\makebox(0,0)[cc]{$W$}}
\put(76.00,115.00){\makebox(0,0)[cc]{$S$}}
\put(91.00,130.00){\makebox(0,0)[cc]{$E$}}
\put(73.00,137.00){\line(2,-1){6.00}}
\put(79.00,137.00){\line(-2,-1){6.00}}
\put(35.00,95.00){\makebox(0,0)[cc]{$\sigma(N,E,S,W)$}}
\put(76.00,95.00){\makebox(0,0)[cc]{$\sigma(0,E,S,W)$}}
\put(115.00,95.00){\makebox(0,0)[cc]{$\sigma(1,E,S,W)$}}
\put(33.00,12.00){\makebox(0,0)[lc]{$\sigma(N_1,\sigma(N_2,E_2,S_2,1),S_1,W_1)$}}
\put(35.00,152.00){\makebox(0,0)[cc]{(a)}}
\put(76.00,152.00){\makebox(0,0)[cc]{(b)}}
\put(115.00,152.00){\makebox(0,0)[cc]{(c)}}
\put(43.00,75.00){\makebox(0,0)[cc]{(d)}}
\put(33.00,45.00){\makebox(0,0)[cc]{\small 1}}
\put(53.00,45.00){\makebox(0,0)[cc]{\small 2}}
\end{picture}
\caption{Samples of application of generating functions.}
\label{fig.2}
\end{figure}
%%%%%%%%%%%%%%%%%%%%%%%%%%%%%%%%%%%%%%%%%%%%%%%%%%%%%%%%%%%%%%%%%%
\newpage
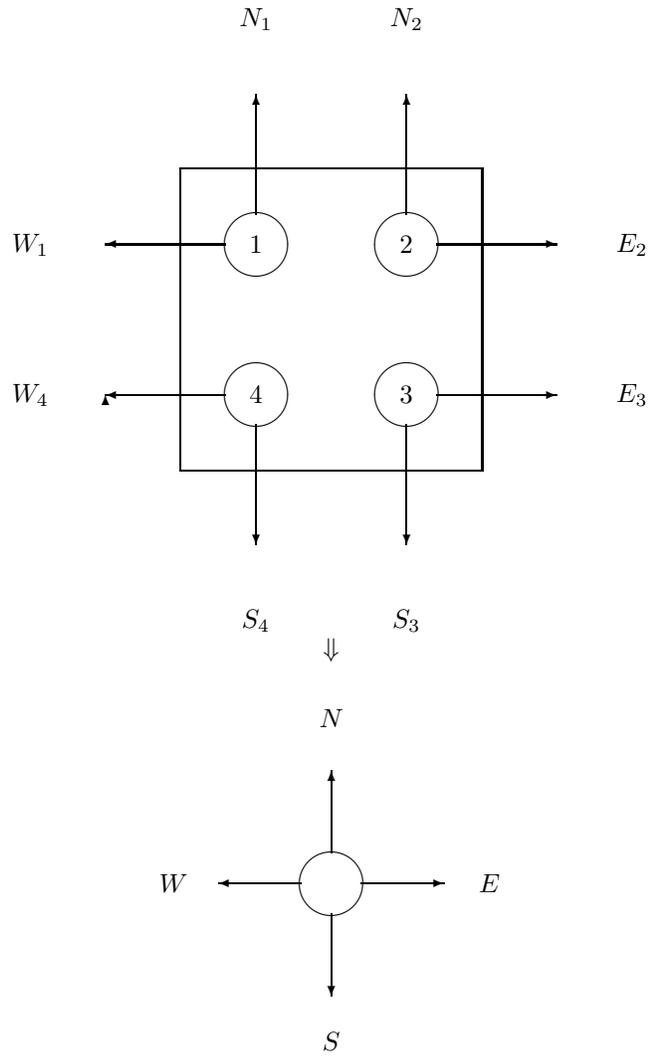
\begin{figure}[h]
\unitlength=1.00mm
\special{em:linewidth 0.4pt}
\linethickness{0.4pt}
\begin{picture}(110.00,150.00)
\put(50.00,90.00){\framebox(40.00,40.00)[cc]{}}
\put(60.00,120.00){\circle{8.25}}
\put(80.00,120.00){\circle{8.00}}
\put(60.00,100.00){\circle{8.00}}
\put(80.00,100.00){\circle{8.00}}
\put(84.00,120.00){\vector(1,0){16.00}}
\put(80.00,124.00){\vector(0,1){16.00}}
\put(60.00,124.00){\vector(0,1){16.00}}
\put(56.00,120.00){\vector(-1,0){16.00}}
\put(80.00,96.00){\vector(0,-1){16.00}}
\put(60.00,96.00){\vector(0,-1){16.00}}
\put(84.00,100.00){\vector(1,0){16.00}}
\put(60.00,120.00){\makebox(0,0)[cc]{$1$}}
\put(80.00,120.00){\makebox(0,0)[cc]{$2$}}
\put(56.00,100.00){\vector(-1,0){16.00}}
\put(40.00,100.00){\vector(0,0){0.00}}
\put(60.00,100.00){\makebox(0,0)[cc]{$4$}}
\put(80.00,100.00){\makebox(0,0)[cc]{$3$}}
\put(60.00,150.00){\makebox(0,0)[cc]{$N_1$}}
\put(80.00,150.00){\makebox(0,0)[cc]{$N_2$}}
\put(110.00,120.00){\makebox(0,0)[cc]{$E_2$}}
\put(110.00,100.00){\makebox(0,0)[cc]{$E_3$}}
\put(80.00,70.00){\makebox(0,0)[cc]{$S_3$}}
\put(60.00,70.00){\makebox(0,0)[cc]{$S_4$}}
\put(30.00,100.00){\makebox(0,0)[cc]{$W_4$}}
\put(30.00,120.00){\makebox(0,0)[cc]{$W_1$}}
\put(70.00,35.00){\circle{8.00}}
\put(70.00,39.00){\vector(0,1){11.00}}
\put(74.00,35.00){\vector(1,0){11.00}}
\put(70.00,31.00){\vector(0,-1){11.00}}
\put(66.00,35.00){\vector(-1,0){11.00}}
\put(70.00,66.00){\makebox(0,0)[cc]{$\Downarrow$}}
\put(70.00,57.00){\makebox(0,0)[cc]{$N$}}
\put(91.00,35.00){\makebox(0,0)[cc]{$E$}}
\put(70.00,14.00){\makebox(0,0)[cc]{$S$}}
\put(49.00,35.00){\makebox(0,0)[cc]{$W$}}
\end{picture}
\caption{Renormalizating the lattice we couple the two bonds
connecting neighbouring blocks into the only bond of the new lattice.
$N_1,N_2 \rightarrow N,~~
E_2,E_3 \rightarrow S,~~
S_3,S_4 \rightarrow W,~~
W_4,W_1 \rightarrow E$.}
\label{fig.3}
\end{figure}
%%%%%%%%%%%%%%%%%%%%%%%%%%%%%%%%%%%%%%%%%%%%%%%%%%%%%%%%%%%%%%%%
\newpage
\begin{figure*}
\unitlength=1mm
\special{em:linewidth 0.5pt}
\linethickness{0.5pt}
\begin{picture}(106.00,120.00)
\thinlines
\put(14.00,98.00){\framebox(20.00,20.00)[cc]{}}
\put(19.00,113.00){\circle{5.66}}
\put(29.00,113.00){\circle{5.66}}
\thicklines
\put(19.00,103.00){\circle{5.66}}
\thinlines
\put(29.00,103.00){\circle{5.66}}
\put(19.00,113.00){\makebox(0,0)[cc]{\scriptsize D}}
\put(19.00,103.00){\makebox(0,0)[cc]{\scriptsize \bf D}}
\put(29.00,103.00){\makebox(0,0)[cc]{\scriptsize X}}
\put(29.00,113.00){\makebox(0,0)[cc]{\scriptsize X}}
\put(19.00,106.00){\vector(0,1){4.00}}
\put(14.00,67.00){\framebox(20.00,20.00)[cc]{}}
\put(19.00,82.00){\circle{5.66}}
\put(29.00,82.00){\circle{5.66}}
\thicklines
\put(19.00,72.00){\circle{5.66}}
\thinlines
\put(29.00,72.00){\circle{5.66}}
\put(19.00,82.00){\makebox(0,0)[cc]{\scriptsize D}}
\put(19.00,72.00){\makebox(0,0)[cc]{\scriptsize \bf D}}
\put(29.00,72.00){\makebox(0,0)[cc]{\scriptsize X}}
\put(29.00,82.00){\makebox(0,0)[cc]{\scriptsize D}}
\put(19.00,75.00){\vector(0,1){4.00}}
\put(49.00,67.00){\framebox(20.00,20.00)[cc]{}}
\put(54.00,82.00){\circle{5.66}}
\put(64.00,82.00){\circle{5.66}}
\thicklines
\put(54.00,72.00){\circle{5.66}}
\thinlines
\put(64.00,72.00){\circle{5.66}}
\put(54.00,82.00){\makebox(0,0)[cc]{\scriptsize D}}
\put(54.00,72.00){\makebox(0,0)[cc]{\scriptsize \bf D}}
\put(64.00,72.00){\makebox(0,0)[cc]{\scriptsize D}}
\put(64.00,82.00){\makebox(0,0)[cc]{\scriptsize X}}
\put(54.00,75.00){\vector(0,1){4.00}}
\put(14.00,36.00){\framebox(20.00,20.00)[cc]{}}
\put(19.00,51.00){\circle{5.66}}
\put(29.00,51.00){\circle{5.66}}
\thicklines
\put(19.00,41.00){\circle{5.66}}
\thinlines
\put(29.00,41.00){\circle{5.66}}
\put(19.00,51.00){\makebox(0,0)[cc]{\scriptsize D}}
\put(19.00,41.00){\makebox(0,0)[cc]{\scriptsize \bf D}}
\put(29.00,41.00){\makebox(0,0)[cc]{\scriptsize C}}
\put(29.00,51.00){\makebox(0,0)[cc]{\scriptsize D}}
\put(19.00,44.00){\vector(0,1){4.00}}
\put(49.00,36.00){\framebox(20.00,20.00)[cc]{}}
\put(54.00,51.00){\circle{5.66}}
\put(64.00,51.00){\circle{5.66}}
\thicklines
\put(54.00,41.00){\circle{5.66}}
\thinlines
\put(64.00,41.00){\circle{5.66}}
\put(54.00,51.00){\makebox(0,0)[cc]{\scriptsize D}}
\put(54.00,41.00){\makebox(0,0)[cc]{\scriptsize \bf D}}
\put(64.00,41.00){\makebox(0,0)[cc]{\scriptsize D}}
\put(64.00,51.00){\makebox(0,0)[cc]{\scriptsize C}}
\put(54.00,44.00){\vector(0,1){4.00}}
\put(14.00,5.00){\framebox(20.00,20.00)[cc]{}}
\put(19.00,20.00){\circle{5.66}}
\put(29.00,20.00){\circle{5.66}}
\thicklines
\put(19.00,10.00){\circle{5.66}}
\thinlines
\put(29.00,10.00){\circle{5.66}}
\put(19.00,20.00){\makebox(0,0)[cc]{\scriptsize D}}
\put(19.00,10.00){\makebox(0,0)[cc]{\scriptsize \bf D}}
\put(29.00,10.00){\makebox(0,0)[cc]{\scriptsize D}}
\put(29.00,20.00){\makebox(0,0)[cc]{\scriptsize D}}
\put(19.00,13.00){\vector(0,1){4.00}}
\put(22.00,82.00){\vector(1,0){4.00}}
\put(57.00,72.00){\vector(1,0){4.00}}
\put(22.00,51.00){\vector(1,0){4.00}}
\put(57.00,41.00){\vector(1,0){4.00}}
\put(29.00,48.00){\vector(0,-1){4.00}}
\put(22.00,41.00){\vector(1,0){4.00}}
\put(57.00,51.00){\vector(1,0){4.00}}
\put(64.00,44.00){\vector(0,1){4.00}}
\put(22.00,20.00){\vector(1,0){4.00}}
\put(24.00,94.00){\makebox(0,0)[cc]{\scriptsize X=A,B,C}}
\put(24.00,63.00){\makebox(0,0)[cc]{\scriptsize X=A,B}}
\put(59.00,63.00){\makebox(0,0)[cc]{\scriptsize X=A,B}}
\put(84.00,67.00){\framebox(20.00,20.00)[cc]{}}
\put(89.00,82.00){\circle{5.66}}
\put(99.00,82.00){\circle{5.66}}
\thicklines
\put(89.00,72.00){\circle{5.66}}
\thinlines
\put(99.00,72.00){\circle{5.66}}
\put(89.00,72.00){\makebox(0,0)[cc]{\scriptsize \bf D}}
\put(99.00,82.00){\makebox(0,0)[cc]{\scriptsize D}}
\put(94.00,63.00){\makebox(0,0)[cc]{\scriptsize X=A,B}}
\put(84.00,36.00){\framebox(20.00,20.00)[cc]{}}
\put(89.00,51.00){\circle{5.66}}
\put(99.00,51.00){\circle{5.66}}
\thicklines
\put(89.00,41.00){\circle{5.66}}
\thinlines
\put(99.00,41.00){\circle{5.66}}
\put(89.00,41.00){\makebox(0,0)[cc]{\scriptsize \bf D}}
\put(99.00,51.00){\makebox(0,0)[cc]{\scriptsize D}}
\put(89.00,44.00){\vector(0,1){4.00}}
\put(92.00,41.00){\vector(1,0){4.00}}
\put(89.00,82.00){\makebox(0,0)[cc]{\scriptsize X}}
\put(99.00,72.00){\makebox(0,0)[cc]{\scriptsize D}}
\put(92.00,72.00){\vector(1,0){4.00}}
\put(99.00,75.00){\vector(0,1){4.00}}
\put(89.00,51.00){\makebox(0,0)[cc]{\scriptsize C}}
\put(99.00,41.00){\makebox(0,0)[cc]{\scriptsize D}}
\put(99.00,44.00){\vector(0,1){4.00}}
\put(96.00,51.00){\vector(-1,0){4.00}}
\put(49.00,5.00){\framebox(20.00,20.00)[cc]{}}
\put(54.00,20.00){\circle{5.66}}
\put(64.00,20.00){\circle{5.66}}
\thicklines
\put(54.00,10.00){\circle{5.66}}
\thinlines
\put(64.00,10.00){\circle{5.66}}
\put(54.00,20.00){\makebox(0,0)[cc]{\scriptsize D}}
\put(54.00,10.00){\makebox(0,0)[cc]{\scriptsize \bf D}}
\put(64.00,10.00){\makebox(0,0)[cc]{\scriptsize D}}
\put(64.00,20.00){\makebox(0,0)[cc]{\scriptsize D}}
\put(54.00,13.00){\vector(0,1){4.00}}
\put(57.00,20.00){\vector(1,0){4.00}}
\put(84.00,5.00){\framebox(20.00,20.00)[cc]{}}
\put(89.00,20.00){\circle{5.66}}
\put(99.00,20.00){\circle{5.66}}
\thicklines
\put(89.00,10.00){\circle{5.66}}
\thinlines
\put(99.00,10.00){\circle{5.66}}
\put(89.00,20.00){\makebox(0,0)[cc]{\scriptsize D}}
\put(89.00,10.00){\makebox(0,0)[cc]{\scriptsize \bf D}}
\put(99.00,10.00){\makebox(0,0)[cc]{\scriptsize D}}
\put(99.00,20.00){\makebox(0,0)[cc]{\scriptsize D}}
\put(29.00,17.00){\vector(0,-1){4.00}}
\put(57.00,10.00){\vector(1,0){4.00}}
\put(64.00,13.00){\vector(0,1){4.00}}
\put(92.00,10.00){\vector(1,0){4.00}}
\put(99.00,13.00){\vector(0,1){4.00}}
\put(96.00,20.00){\vector(-1,0){4.00}}
\put(12.00,3.00){\dashbox{1.00}(94.00,24.00)[cc]{}}
\put(0.00,108.00){\makebox(0,0)[lc]{(a)}}
\put(0.00,77.00){\makebox(0,0)[lc]{(b)}}
\put(0.00,46.00){\makebox(0,0)[lc]{(c)}}
\put(0.00,15.00){\makebox(0,0)[lc]{(d)}}
\end{picture}
\caption{We show the four different types of the blocks for the cemichal
reaction model and some relaxation schemes spanning them.
The other schemes can be obtained from these figures
by rotations. It is convenient for calculations to subdivide the relaxation
processes in the block (d) into three parts shown in the dashed box.}
\label{fig.4}
\end{figure*}
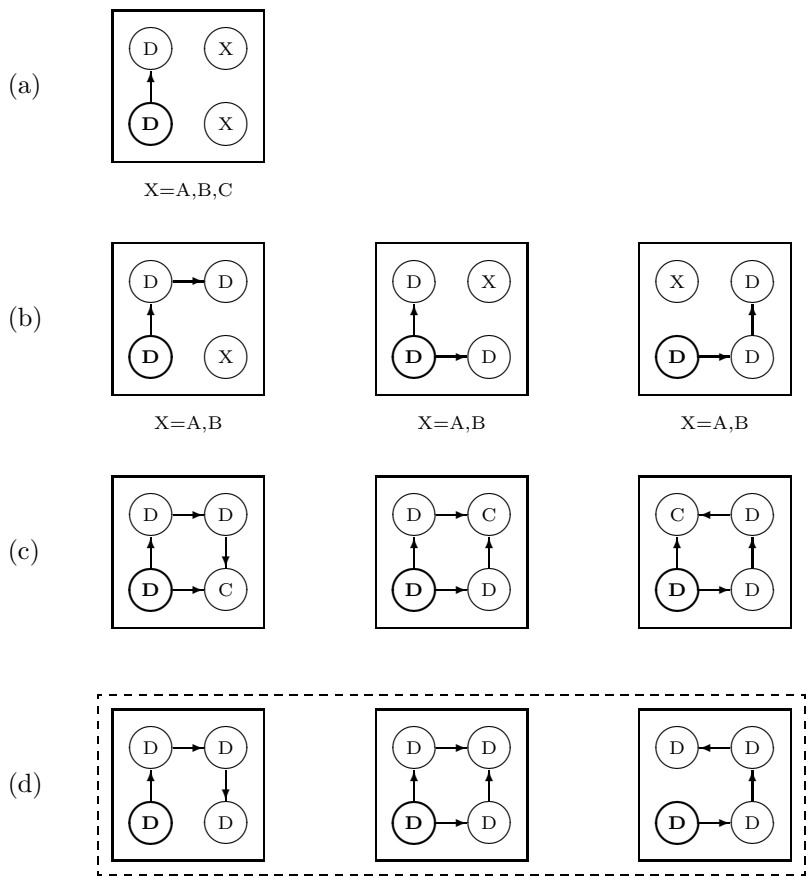
%%%%%%%%%%%%%%%%%%%%%%%%%%%%%%%%%%%%%%%%%%%%%%%%%%%%%%%%%%%%%%%%%%
\newpage
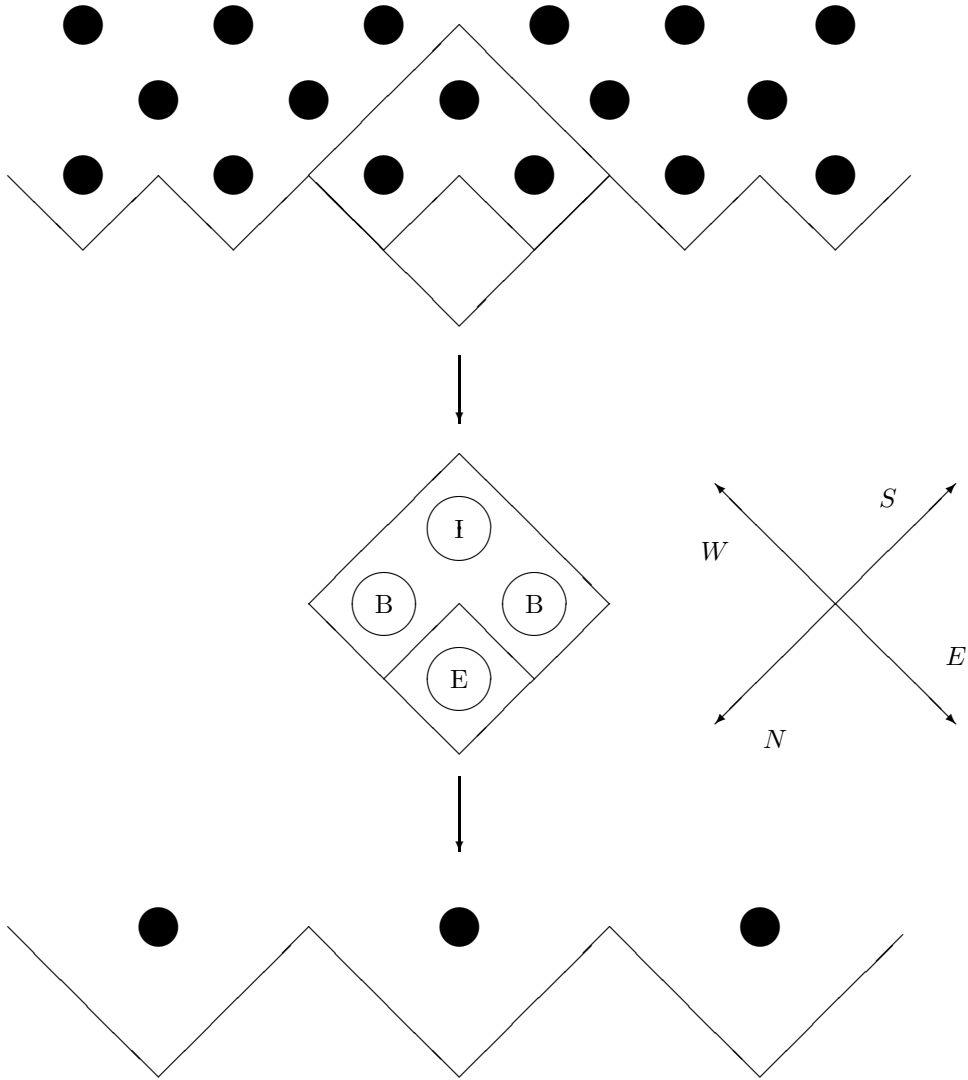
\begin{figure}[h]
\unitlength=1mm
\special{em:linewidth 0.4pt}
\linethickness{0.4pt}
\begin{picture}(146.00,162.60)
\put(10.00,120.00){\line(1,-1){10.00}}
\put(20.00,110.00){\line(1,1){10.00}}
\put(30.00,120.00){\line(1,-1){10.00}}
\put(40.00,110.00){\line(1,1){10.00}}
\put(50.00,120.00){\line(1,-1){10.00}}
\put(60.00,110.00){\line(1,1){10.00}}
\put(70.00,120.00){\line(1,-1){10.00}}
\put(80.00,110.00){\line(1,1){10.00}}
\put(90.00,120.00){\line(1,-1){10.00}}
\put(100.00,110.00){\line(1,1){10.00}}
\put(110.00,120.00){\line(1,-1){10.00}}
\put(120.00,110.00){\line(1,1){10.00}}
\put(20.00,120.00){\circle*{5.20}}
\put(40.00,120.00){\circle*{5.20}}
\put(60.00,120.00){\circle*{5.20}}
\put(80.00,120.00){\circle*{5.20}}
\put(100.00,120.00){\circle*{5.20}}
\put(120.00,120.00){\circle*{5.20}}
\put(30.00,130.00){\circle*{5.20}}
\put(50.00,130.00){\circle*{5.20}}
\put(70.00,130.00){\circle*{5.20}}
\put(90.00,130.00){\circle*{0.00}}
\put(90.00,130.00){\circle*{5.20}}
\put(111.00,130.00){\circle*{5.20}}
\put(20.00,140.00){\circle*{5.20}}
\put(40.00,140.00){\circle*{5.20}}
\put(60.00,140.00){\circle*{5.20}}
\put(82.00,140.00){\circle*{5.20}}
\put(100.00,140.00){\circle*{5.20}}
\put(120.00,140.00){\circle*{5.20}}
\put(50.00,120.00){\line(1,1){20.00}}
\put(70.00,140.00){\line(1,-1){20.00}}
\put(90.00,120.00){\line(-1,-1){20.00}}
\put(70.00,100.00){\line(-1,1){19.00}}
\put(70.00,43.00){\line(-1,1){20.00}}
\put(50.00,63.00){\line(1,1){20.00}}
\put(70.00,83.00){\line(1,-1){20.00}}
\put(90.00,63.00){\line(-1,-1){20.00}}
\put(60.00,53.00){\line(1,1){10.00}}
\put(70.00,63.00){\line(1,-1){10.00}}
\put(70.00,53.00){\circle{8.00}}
\put(60.00,63.00){\circle{8.00}}
\put(70.00,73.00){\circle{0.00}}
\put(70.00,73.00){\circle{8.00}}
\put(80.00,63.00){\circle{8.25}}
\put(60.00,63.00){\makebox(0,0)[cc]{B}}
\put(80.00,63.00){\makebox(0,0)[cc]{B}}
\put(70.00,73.00){\makebox(0,0)[cc]{I}}
\put(70.00,53.00){\makebox(0,0)[cc]{E}}
\put(70.00,96.00){\vector(0,-1){9.00}}
\put(120.00,63.00){\vector(1,-1){16.00}}
\put(120.00,63.00){\vector(-1,-1){16.00}}
\put(120.00,63.00){\vector(1,1){16.00}}
\put(120.00,63.00){\vector(-1,1){16.00}}
\put(112.00,45.00){\makebox(0,0)[cc]{$N$}}
\put(136.00,56.00){\makebox(0,0)[cc]{$E$}}
\put(127.00,77.00){\makebox(0,0)[cc]{$S$}}
\put(104.00,70.00){\makebox(0,0)[cc]{$W$}}
\put(10.00,20.00){\line(1,-1){20.00}}
\put(30.00,0.00){\line(1,1){20.00}}
\put(50.00,20.00){\line(1,-1){20.00}}
\put(70.00,0.00){\line(1,1){20.00}}
\put(90.00,20.00){\line(1,-1){20.00}}
\put(110.00,0.00){\line(1,1){19.00}}
\put(30.00,20.00){\circle*{5.20}}
\put(70.00,20.00){\circle*{5.20}}
\put(110.00,20.00){\circle*{5.20}}
\put(70.00,40.00){\vector(0,-1){10.00}}
\end{picture}
\vspace{1cm}
\caption{The procedure of cell-to-site transformation
at the open boundary of the lattice. One internal cell (I),
two boundary cells (B) and one auxiliary external cell (E)
build the boundary cell at the next scale $2b$.}
\label{fig.5}
\end{figure}


\begin{thebibliography}{99}

\bibitem{domb}
C. Domb and M. S. Green (eds),
{\em ``Phase Transition and Critical Phenomena''},
vols 1-6, Academic Press (London, 1972-76);
C. Domb and J. L. Lebowitz (eds),
{\em ``Phase Transition and Critical Phenomena''},
vols 7-17 , Academic Press (London, 1983-95).
\bibitem{katz}
S. Katz, J. L. Lebowitz and H. Spohn,
Phys. Rev.B {\bf 28}, 1655 (1983);
J.Stat. Phys. {\bf 34}, 497 (1984).
\bibitem{zia}
B. Schmittmann and R. K. Zia,
in ref.\cite{domb} vol.17 (1995).
\bibitem{soc}
For a review see e.g.:
P. Bak and M. Creutz,
in ''{\em Fractals and Disordered Systems}''
vol.II, ed. by A. Bunde and S. Havlin,
Springer Verlag, Heidelberg (1993).
\bibitem{vic}
T. Vicsek, {\it Fractal Growth Phenomena},
World Scientific, Singapore (1992).
\bibitem{bb}
B. B. Mandelbrot,
{\em ``The Fractal Geometry of Nature''},
(Freeman and Company, New York, 1983).
\bibitem{bak2}
P. Bak, C. Tang and K. Wiesenfeld,
Phys.Rev.Lett. {\bf 59}, 381 (1987);
Phys. Rev.A {\bf 38}, 364 (1988);
G. Grinstein in ''{\em Scale Invariance, Interfaces and Non-Equilibrium 
Dynamics}'', ed. by A. McKane et al, 
NATO Advanced Study Institute, Series B:
Physics Vol. 344 (Plenum, New York, 1995).
\bibitem{bak}
P. Bak, K. Chen and C. Tang,
Phys. Lett. A {\bf 147}, 297 (1990).
\bibitem{ds} B. Drossel and F. Schwabl,
Phys. Rev. Lett. {\bf 69}
1629 (1992).
\bibitem{vzmf}
A.Vespignani and S.Zapperi, Phys. Rev. Lett. {\bf 78}
4793 (1997); Phys. Rev.E {\bf 57}, xxx (1998);
R. Dickman, A. Vespignani and S. Zapperi, 
Phys. Rev.E {\bf 57}, xxx (1998).
\bibitem{pvz}
L. Pietronero, A. Vespignani and S. Zapperi,
Phys. Rev. Lett. {\bf 72}, 1690 (1994);
A. Vespignani, S. Zapperi and L.Pietronero,
Phys. Rev E {\bf 51}, 1711 (1995).
\bibitem{lpvz}
V. Loreto, L. Pietronero, A. Vespignani and S. Zapperi,
Phys. Rev. Lett. {\bf 75}, 465 (1995).
\bibitem{ddrg}
	A. Vespignani, S. Zapperi and V. Loreto,
        Phys. Rev. Lett. {\bf 77}, 4560 (1996);
	J. Stat. Phys. {\bf 88}, 47 (1997).
\bibitem{iva}
	E. V. Ivashkevich,
	Phys. Rev. Lett. {\bf 76}, 3368 (1996).
\bibitem{wies}
	J. Hasty and K. Wiesenfeld,
	J. Stat. Phys., {\bf 86}, 1179 (1997).
\bibitem{oliv}
	T. Tom\'e and M.J. de Oliveira,
	Phys. Rev. E {\bf 55}, 4000 (1997);
	M.J. de Oliveira and J.E. Satulovsky,
	Phys. Rev. E {\bf 55}, 6377 (1997).
\bibitem{zhang}
        Y. C. Zhang,
        Phys. Rev. Lett. {\bf 63}, 470 (1989);
        L. Pietronero, P. Tartaglia and Y. C. Zhang,
        Physica A {\bf 173}, 129 (1991).
\bibitem{manna}
        S. S. Manna, J. Phys. A {\bf 24}, L363 (1991).
\bibitem{grasma}
        P. Grassberger and S. S. Manna,
        J. Phys. (France) {\bf 51}, 1077 (1990).
\bibitem{stella}
        A. L. Stella, C. Tebaldi and G. Caldarelli,
        Phys. Rev. E {\bf 52}, 72 (1995).
\bibitem{lubeck}
	S. L\"ubeck and K. D. Usadel,
        Phys. Rev. E {\bf 55}, 4095 (1997);
        S. L\"ubeck, Phys. Rev. E {\bf 56}, 1590 (1997).
\bibitem{D}
	D. Dhar, Phys.Rev.Lett. {\bf 64}, 1613 (1990).
\bibitem{MDheight}
	S.N. Majumdar and D.Dhar, J.Phys. {\bf A 24}, L357 (1991).
\bibitem{P}
	V.B. Priezzhev, J. Stat. Phys. {\bf 74} 955, (1994).
\bibitem{I}
	E.V. Ivashkevich, J. Phys. {\bf A 27} 3643 (1994).
\bibitem{IKP}
	E.V. Ivashkevich, D.V. Ktitarev and V.B. Priezzhev,
	J. Phys. A {\bf 27}, L585 (1994).
\bibitem{wave}
       V.B.~Priezzhev, D.V. Ktitarev and E. V. Ivashkevich,
       Physica A 209 (1994) 347
\bibitem{PKI}
	V.B.~Priezzhev, D.V. Ktitarev and E. V. Ivashkevich,
	Phys. Rev. Lett. {\bf 76}, 2093 (1996).
\bibitem{nv}
	T. Niemeijer and J. M. J. van Leeuwen,
	in ref.\cite{domb} vol.6 (1976).
\bibitem{kei}
	J. Keizer,
	{\em ``Statistical Thermodynamics of Nonequilibrium Processes''},
	Springer Verlag (New York, 1987).
\bibitem{dick}
	R. Dickman,
	Phys. Rev. {\bf A 38}, 2588 (1988).
\bibitem{ben}
	A. Ben-Hur and O.Biham,
	Phys. Rev. E {\bf 53}, R1317 (1996).
\bibitem{cardy-84}
        Cardy J L 1987 {\it Phase Transitions and Critical Phenomena} vol {\bf 11},
        eds C Domb and J L Lebowitz (London: Academic Press) p 55
\end{thebibliography}
\end{document}